\documentclass[nojss,article]{jss}\usepackage[]{graphicx}\usepackage[]{color}
\makeatletter
\def\maxwidth{ %
  \ifdim\Gin@nat@width>\linewidth
    \linewidth
  \else
    \Gin@nat@width
  \fi
}
\makeatother

\definecolor{fgcolor}{rgb}{0.345, 0.345, 0.345}
\newcommand{\hlnum}[1]{\textcolor[rgb]{0.686,0.059,0.569}{#1}}%
\newcommand{\hlstr}[1]{\textcolor[rgb]{0.192,0.494,0.8}{#1}}%
\newcommand{\hlcom}[1]{\textcolor[rgb]{0.678,0.584,0.686}{\textit{#1}}}%
\newcommand{\hlopt}[1]{\textcolor[rgb]{0,0,0}{#1}}%
\newcommand{\hlstd}[1]{\textcolor[rgb]{0.345,0.345,0.345}{#1}}%
\newcommand{\hlkwb}[1]{\textcolor[rgb]{0.69,0.353,0.396}{#1}}%
\newcommand{\hlkwc}[1]{\textcolor[rgb]{0.333,0.667,0.333}{#1}}%
\newcommand{\hlkwd}[1]{\textcolor[rgb]{0.737,0.353,0.396}{\textbf{#1}}}%

\usepackage{framed}
\makeatletter
\newenvironment{kframe}{%
 \def\at@end@of@kframe{}%
 \ifinner\ifhmode%
  \def\at@end@of@kframe{\end{minipage}}%
  \begin{minipage}{\columnwidth}%
 \fi\fi%
 \def\FrameCommand##1{\hskip\@totalleftmargin \hskip-\fboxsep
 \colorbox{shadecolor}{##1}\hskip-\fboxsep
     \hskip-\linewidth \hskip-\@totalleftmargin \hskip\columnwidth}%
 \MakeFramed {\advance\hsize-\width
   \@totalleftmargin\z@ \linewidth\hsize
   \@setminipage}}%
 {\par\unskip\endMakeFramed%
 \at@end@of@kframe}
\makeatother

\definecolor{shadecolor}{rgb}{.97, .97, .97}
\definecolor{messagecolor}{rgb}{0, 0, 0}
\definecolor{warningcolor}{rgb}{1, 0, 1}
\definecolor{errorcolor}{rgb}{1, 0, 0}
\newenvironment{knitrout}{}{} 

\usepackage{alltt}

\usepackage{amsmath}
\usepackage{array}

\newcolumntype{L}[1]{>{\raggedright\let\newline\\\arraybackslash\hspace{0pt}}m{#1}}

\author{Satu Helske\\ Link\"oping University, Sweden\\ University of Oxford, UK\\ University of Jyv{\"a}skyl{\"a}, Finland \And
        Jouni Helske\\ Link\"oping University, Sweden\\ University of Jyv{\"a}skyl{\"a}, Finland}
\title{Mixture Hidden Markov Models for Sequence Data: The \pkg{seqHMM} Package in \proglang{R}}

\Plainauthor{Satu Helske, Jouni Helske} 
\Plaintitle{Mixture Hidden Markov Models for Sequence Data: The seqHMM Package in R} 
\Shorttitle{\pkg{seqHMM}: Mixture Hidden Markov Models for Sequence Data} 

\Abstract{
Sequence analysis is being more and more widely used for the analysis of social sequences and other multivariate categorical time series data. However, it is often complex to describe, visualize, and compare large sequence data, especially when there are multiple parallel sequences per subject. Hidden (latent) Markov models (HMMs) are able to detect underlying latent structures and they can be used in various longitudinal settings: to account for measurement error, to detect unobservable states, or to compress information across several types of observations. Extending to mixture hidden Markov models (MHMMs) allows clustering data into homogeneous subsets, with or without external covariates.

The \pkg{seqHMM} package in \proglang{R} is designed for the efficient modeling of sequences and other categorical time series data containing one or multiple subjects with one or multiple interdependent sequences using HMMs and MHMMs. Also other restricted variants of the MHMM can be fitted, e.g., latent class models, Markov models, mixture Markov models, or even ordinary multinomial regression models with suitable parameterization of the HMM.

Good graphical presentations of data and models are useful during the whole analysis process from the first glimpse at the data to model fitting and presentation of results. The package provides easy options for plotting parallel sequence data, and proposes visualizing HMMs as directed graphs.
}
\Plainkeywords{multichannel sequences, categorical time series, visualizing sequence data, visualizing models, latent Markov models, latent class models, R} 

\Address{
  Satu Helske\\
  Institute for Analytical Sociology \\
  Link\"oping University \\
  SE-60174 Norrk\"oping\\
  Sweden \\
  E-mail: \email{satu.helske@liu.se}\\
  
  Jouni Helske\\
  Department of Science and Technology \\
  Link\"oping University\\
  SE-60174 Norrk\"oping\\
  Sweden\\
  E-mail: \email{Jouni.Helske@liu.se}
}

\IfFileExists{upquote.sty}{\usepackage{upquote}}{}
\begin{document}

Vignette based on the corresponding paper at \emph{Journal of Statistical Software} \citep{JSS2019}.

\section{Introduction}

Social sequence analysis is being more and more widely used for the analysis of longitudinal data consisting of multiple independent subjects with one or multiple interdependent sequences (channels). Sequence analysis is used for computing the (dis)similarities of sequences, and often the goal is to find patterns in data using cluster analysis. However, describing, visualizing, and comparing large sequence data is often complex, especially in the case of multiple channels. Hidden (latent) Markov models (HMMs) can be used to compress and visualize information in such data. These models are able to detect underlying latent structures. Extending to mixture hidden Markov models (MHMMs) allows clustering via latent classes, possibly with additional covariate information. One of the major benefits of using hidden Markov modeling is that all stages of analysis are performed, evaluated, and compared in a probabilistic framework.

The \pkg{seqHMM} \citep{JSS2019} package for \proglang{R} \citep{Rcore} is designed for modeling sequence data and other categorical time series with one or multiple subjects and one or multiple channels using HMMs and MHMMs. The package provides functions for the estimation and inference of models, as well as functions for the easy visualization of multichannel sequences and HMMs. Even though the package was originally developed for researchers familiar with social sequence analysis and the examples are related to life course, knowledge on sequence analysis or social sciences is not necessary for the usage of \pkg{seqHMM}. The package is available on Comprehensive R Archive Repository (CRAN) and easily installed via \code{install.packages("seqHMM")}. Development versions can be obtained from GitHub\footnote{\url{https://github.com/helske/seqHMM}}.

There are also other \proglang{R} packages in CRAN for HMM analysis of categorical data. The \pkg{HMM} package \citep{Himmelmann2010} is a compact package designed for fitting an HMM for a single observation sequence. The \pkg{hmm.discnp} package \citep{Turner2014} can handle multiple observation sequences with possibly varying lengths. For modeling continuous-time processes as hidden Markov models, the \pkg{msm} package \citep{Jackson2011} is available. Both \pkg{hmm.discnp} and \pkg{msm} support only single-channel observations. The \pkg{depmixS4} package \citep{Visser2010} is able to fit HMMs for multiple interdependent time series (with continuous or categorical values), but for one subject only. In the \pkg{msm} and \pkg{depmixS4} packages, covariates can be added for initial and transition probabilities. The \pkg{mhsmm} package \citep{OConnell2011} allows modeling of multiple sequences using hidden Markov and semi-Markov models. There are no ready-made options for modeling categorical data, but users can write their own extensions for arbitrary distributions. The \pkg{LMest} package \citep{Bartolucci2015} is aimed to panel data with a large number of subjects and a small number of time points. It can be used for hidden Markov modeling of multivariate and multichannel categorical data, using covariates in emission and transition processes. \pkg{LMest} also supports mixed latent Markov models, where the latent process is allowed to vary in different latent subpopulations. This differs from mixture hidden Markov models used in \pkg{seqHMM}, where also the emission probabilities vary between groups. The \pkg{seqHMM} package also supports covariates in explaining group memberships. A drawback in the \pkg{LMest} package is that the user cannot define initial values or zero constraints for model parameters, and thus important special cases such as left-to-right models cannot be used.

We start with describing data and methods: a short introduction to sequence data and sequence analysis, then the theory of hidden Markov models for such data, an expansion to mixture hidden Markov models and a glance at some special cases, and then some propositions on visualizing multichannel sequence data and hidden Markov models. After the theoretic part we take a look at features of the \pkg{seqHMM} package and at the end show an example on using the package for the analysis of life course data. The appendix shows the list of notations.

\newpage
\section{Methods}
\subsection{Sequences and sequence analysis}
\label{sec:seqdata}

By the term \emph{sequence} we refer to an ordered set of categorical states. It can be a time series, such as a career trajectory or residential history, or any other series with ordered categorical observations, e.g., a DNA sequence or a structure of a story. Typically, \emph{sequence data} consist of multiple independent subjects (\emph{multivariate} data). Sometimes there are also multiple interdependent sequences per subject, often referred to as \emph{multichannel} or multidimensional sequence data.

As an example we use the \code{biofam} data available in the \pkg{TraMineR} package \citep{Gabadinho2011}. It is a sample of 2000 individuals born in 1909--1972, constructed from the Swiss Household Panel survey in 2002 \citep{Mueller2007}. The data set contains sequences of annual family life statuses from age 15 to 30. Eight observed states are defined from the combination of five basic states: living with parents, left home, married, having children, and divorced. To show a more complex example, we split the original data into three separate \emph{channels} representing different life domains: marriage, parenthood, and residence. The data for each individual now includes three parallel sequences constituting of two or three \emph{states} each: single/married/divorced, childless/parent, and living with parents / having left home.

\emph{Sequence analysis} (SA), as defined in the social science framework, is a model-free data-driven approach to the analysis of successions of states. The approach has roots in bioinformatics and computer science \citep[see e.g.][]{Durbin1998}, but during the past few decades SA has also become more common in other disciplines for the analysis of longitudinal data. In social sciences SA has been used increasingly often and is now ``central to the life-course perspective'' \citep{Blanchard2014}. 

SA is used for computing (dis)similarities of sequences. The most well-known method is optimal matching \citep{McVicar2002}, but several alternatives exist \citep[see e.g.][]{Aisenbrey2010, Elzinga2014, Gauthier2009, Halpin2010, Hollister2009, Lesnard2010}. Also a method for analyzing multichannel data has been developed \citep{Gauthier2010}. Often the goal in SA is to find typical and atypical patterns in trajectories using cluster analysis, but any approach suitable for compressing information on the dissimilarities can be used. The data are usually presented also graphically in some way. So far the \pkg{TraMineR} package has been the most extensive and frequently used software for social sequence analysis.

\subsection{Hidden Markov models}\label{sec:fitHMM}

In the context of hidden Markov models, sequence data consists of \emph{observed states}, which are regarded as probabilistic functions of \emph{hidden states}. Hidden states cannot be observed directly, but only through the sequence(s) of observations, since they emit the observations on varying probabilities. A discrete first order hidden Markov model for a single sequence is characterized by the following:

\begin{itemize}
\item \emph{Observed state sequence} $\textbf{y}=(y_1, y_2, \ldots, y_T)$
  with observed states $m\in\{1, \ldots, M\}$.

\item \emph{Hidden state sequence} $\textbf{z}=(z_1, z_2,\ldots, z_T)$
with hidden states $s\in\{1, \ldots, S\}$.

\item \emph{Initial probability vector} $\pi=\{\pi_s\}$ of length $S$, where $\pi_s$ is the probability of starting from the hidden state $s$:
  			\[\pi_s = P(z_1=s);\quad s\in \{1, \ldots, S\}.
				\]
				
\item \emph{Transition matrix} $A = \{a_{sr}\}$ of size $S \times S$, where $a_{sr}$ is the probability of moving from the hidden state $s$ at time $t-1$ to the hidden state $r$ at time $t$:
        \[a_{sr}	= P(z_t=r|z_{t-1}=s);\quad s,r \in\{1, \ldots, S\}.
				\]
        We only consider homogeneous HMMs, where the transition probabilities $a_{sr}$ are constant over time.

\item \emph{Emission matrix} $B=\{b_s(m)\}$ of size $S \times M$, where $b_s(m)$ is the probability of the hidden state $s$ emitting the observed state $m$:
  			\[b_{s}(m)=P(y_t=m|z_t=s); \quad s\in\{1, \ldots, S\}, m\in\{1, \ldots, M\}.
				\]

  \end{itemize}
The (first order) Markov assumption states that the hidden state transition probability at time $t$ only depends on the hidden state at the previous time point $t-1$:
\begin{equation}
P(z_t|z_{t-1},\ldots,z_1)=P(z_t|z_{t-1}).
\end{equation}
Also, the observation at time $t$ is only dependent on the current hidden state, not on previous hidden states or observations:
\begin{equation}
P(y_t|y_{t-1},\ldots,y_1,z_t,\ldots,z_1)=P(y_t|z_t).
\end{equation}
For a more detailed description of hidden Markov models, see e.g., \citet{Rabiner1989}, \citet{MacDonald1997}, and \citet{Durbin1998}.

\subsubsection{HMM for multiple sequences}

We can also fit the same HMM for multiple subjects; instead of one observed sequence $\textbf{y}$ we have $N$ sequences as $Y = (\textbf{y}_1, \ldots, \textbf{y}_N)^{\top}$, where the observations $\textbf{y}_i=(y_{i1},\ldots, y_{iT})$ of each subject $i$ take values in the observed state space. Observed sequences are assumed to be mutually independent given the hidden states. The observations are assumed to be generated by the same model, but each subject has its own hidden state sequence.

\subsubsection{HMM for multichannel sequences}

In the case of multichannel sequence data, such as the example described in Section \ref{sec:seqdata}, for each subject $i$ there are $C$ parallel sequences. Observations are now of the form $y_{itc}, \ i=1,\ldots,N; \quad t=1\ldots,T; \quad c=1\ldots,C$, so that our complete data is $Y = \{Y^1, \ldots, Y^C\}$. In \pkg{seqHMM}, multichannel data are handled as a list of $C$ data frames of size $N\times T$. We also define $Y_i$ as all the observations corresponding to subject $i$.

We apply the same latent structure for all channels. In such a case the model has one transition matrix $A$ but several emission matrices $B_1,\ldots,B_C$, one for each channel. We assume that the observed states in different channels at a given time point $t$ are independent of each other given the hidden state at $t$, i.e., $P(\textbf{y}_{it}|z_{it})=P(y_{it1}|z_{it})\cdots P(y_{itC}|z_{it})$.

Sometimes the independence assumption does not seem theoretically plausible. For example, even conditioning on a hidden state representing a general life stage, are marital status and parenthood truly independent? On the other hand, given a person's religious views, could their opinions on abortion and gay marriage be though as independent?

If the goal is to use hidden Markov models for prediction or simulating new sequence data, the analyst should carefully check the validity of independence assumptions. However, if the goal is merely to describe structures and compress information, it can be useful to accept the independence assumption even though it is not completely reasonable in a theoretical sense. When using multichannel sequences, the number of observed states is smaller, which leads to a more parsimonious representation of the model and easier inference of the phenomenon. Also due to the decreased number of observed states, the number of parameters of the model is decreased leading to the improved computational efficiency of model estimation.

The multichannel approach is particularly useful if some of the channels are only partially observed; combining missing and non-missing information into one observation is usually problematic. One would have to decide whether such observations are coded completely missing, which is simple but loses information, or whether all possible combinations of missing and non-missing states are included, which grows the state space larger and makes the interpretation of the model more difficult. In the multichannel approach the data can be used as it is.

\subsubsection{Missing data}

Missing observations are handled straightforwardly in the context of HMMs. When observation $y_{itc}$ is missing, we gain no additional information regarding hidden states. In such a case, we set the emission probability $b_{s}(y_{itc})=1$ for all $s \in {1,\ldots,S}$. Sequences with varying lengths are handled by setting missing values before and/or after the observed states.

\subsubsection{Log-likelihood and parameter estimation}
\label{sec:estimation}

The unknown transition, emission and initial probabilities are commonly estimated via maximum likelihood. The log-likelihood of the parameters $\mathcal{M} = \{\pi, A, B_1, \ldots, B_C\}$ for the HMM is written as
\begin{equation}
\label{eq:L1}
 \log L = \sum_{i=1}^N \log P\left(Y_i|\mathcal{M}\right),
\end{equation}
where $Y_i$ are the observed sequences in channels $c=1,\ldots,C$ for subject $i$. The probability of the observation sequence of subject $i$ given the model parameters is
  \begin{equation}\label{eq:obsprob}
  \begin{aligned}
	P(Y_i|\mathcal{M})	&= \sum_{\text{all } z} P\left(Y_i|z,\mathcal{M}\right) P\left(z|\mathcal{M}\right)\\
	&= \sum_{\text{all } z}  P(z_1|\mathcal{M}) P(\textbf{y}_{i1}|z_1, \mathcal{M})	\prod_{t=2}^T P(z_t|z_{t-1},\mathcal{M}) P(\textbf{y}_{it}|z_t,\mathcal{M}) \\
	&= \sum_{\text{all } z} \pi_{z_1} b_{z_1}(y_{i11})\cdots b_{z_1}(y_{i1C})
	  \prod_{t=2}^T \left[ a_{z_{t-1}z_t} b_{z_t}(y_{it1})\cdots b_{z_t}(y_{itC}) \right],
	\end{aligned}
		    \end{equation}
where the hidden state sequences $z=(z_1,\ldots, z_T)$ take all possible combinations of values in the hidden state space $\{1,\ldots,S\}$ and where $\textbf{y}_{it}$ are the observations of subject $i$ at $t$ in channels $1,\ldots,C$; $\pi_{z_1}$ is the initial probability of the hidden state at time $t=1$ in sequence $z$; $a_{z_{t-1}z_t}$ is the transition probability from the hidden state at time $t-1$ to the hidden state at $t$; and $b_{z_t}(y_{itc})$ is the probability that the hidden state of subject $i$ at time $t$ emits the observed state at $t$ in channel $c$.

For direct numerical maximization (DNM) of the log-likelihood, any general-purpose optimization routines such as BFGS or Nelder--Mead can be used (with suitable reparameterizations). Another common estimation method is the expectation--maximization (EM) algorithm, also known as the Baum--Welch algorithm in the HMM context. The EM algorithm rapidly converges close to a local optimum, but compared to DNM, the converge speed is often slow near the optimum.

The probability \eqref{eq:obsprob} is efficiently calculated using the forward part of the \emph{for\-ward--back\-ward algorithm} \citep{Baum1966, Rabiner1989}. The backward part of the algorithm is needed for the EM algorithm, as well as for the computation of analytic gradients for derivative based optimization routines. For more information on the algorithms, see a supplementary vignette on CRAN \citep{Helske2017c}.

The estimation process starts by giving initial values to the estimates. Good starting values are needed for finding the optimal solution in a reasonable time. In order to reduce the risk of being trapped in a poor local maximum, a large number of initial values should be tested.

\subsubsection{Inference on hidden states}
\label{sec:mpp}

Given our model and observed sequences, we can make several interesting inferences regarding the hidden states. Forward probabilities $\alpha_{it}(s)$ \citep{Rabiner1989} are defined as the joint probability of hidden state $s$ at time $t$ and the observation sequences
$\textbf{y}_{i1}, \ldots, \textbf{y}_{it}$ given the model $\mathcal{M}$, whereas backward probabilities $\beta_{it}(s)$ are defined as the joint probability of hidden state $s$ at time $t$ and the observation sequences $\textbf{y}_{i(t+1)}, \ldots, \textbf{y}_{iT}$ given the model $\mathcal{M}$.

From forward and backward probabilities we can compute the \emph{posterior probabilities} of states, which give the probability of being in each hidden state at each time point, given the observed sequences of subject $i$. These are defined as
\begin{equation}
P(z_{it} = s | Y_i, \mathcal{M}) = \frac{\alpha_{it} \beta_{it}}{P\left(Y_i|\mathcal{M}\right)}.
\end{equation}
Posterior probabilities can be used to find the locally most probable hidden state at each time point, but the resulting sequence is not necessarily globally optimal. To find the single best hidden state sequence $\hat{z}_i(Y_i)=\hat{z}_{i1},  \hat{z}_{i2}, \ldots, \hat{z}_{iT}$ for subject $i$, we maximize $P(z|Y_i,\mathcal{M})$ or, equivalently, $P(z, Y_i|\mathcal{M})$. A dynamic programming method, the \emph{Viterbi algorithm} \citep{Rabiner1989}, is used for solving the problem.

\subsubsection{Model comparison}

Models with the same number of parameters can be compared with the log-likelihood. For choosing between models with a different number of hidden states, we need to take account of the number of parameters. We define the Bayesian information criterion (BIC) as
	\begin{equation}
	BIC=-2\log(L_d)+p\log \left(\sum_{i = 1}^N \sum_{t = 1}^T \frac{1}{C} \sum_{c = 1}^C \text{I}(y_{itc} \text{ observed}) \right),
	\end{equation}
where $L_d$ is computed using Equation~\ref{eq:L1}, $p$ is the number of estimated parameters, I is the indicator function, and the summation in the logarithm is the size of the data. If data are completely observed, the summation is simplified to $N \times T$. Missing observations in multichannel data may lead to non-integer data size.

\subsection{Clustering by mixture hidden Markov models}
\label{sec:MHMM}

There are many approaches for finding and describing clusters or latent classes when working with HMMs. A simple option is to group sequences beforehand (e.g., using sequence analysis and a clustering method), after which one HMM is fitted for each cluster. This approach is simple in terms of HMMs. Models with a different number of hidden states and initial values are explored and compared one cluster at a time. HMMs are used for compressing information and comparing different clustering solutions, e.g., finding the best number of clusters. The problem with this solution is that it is, of course, very sensitive to the original clustering and the estimated HMMs might not be well suited for borderline cases.

Instead of fixing sequences into clusters, it is possible to fit one model for the whole data and determine clustering during modeling. Now sequences are not in fixed clusters but get assigned to clusters with certain probabilities during the modeling process. In this section we expand the idea of HMMs to mixture hidden Markov models (MHMMs). This approach was formulated by \citet{Pol1990} as a mixed Markov latent class model and later generalized to include time-constant and time-varying covariates by \citet{Vermunt2008} (who named the resulting model as mixture latent Markov model, MLMM). The MHMM presented here is a variant of MLMM where only time-constant covariates are allowed. Time-constant covariates deal with unobserved heterogeneity and they are used for predicting cluster memberships of subjects.

\subsubsection{Mixture hidden Markov model}

Assume that we have a set of HMMs $\mathcal{M} = \{\mathcal{M}^1,\ldots, \mathcal{M}^K\}$, where $\mathcal{M}^k = \{\pi^k, A^k, B_1^k, \ldots, B_C^k\}$ for submodels $k=1, \ldots, K$. For each subject $Y_i$, denote $P(\mathcal{M}^k) = w_k$ as the prior probability that the observation sequences of a subject follow the submodel $\mathcal{M}^k$. Now the log-likelihood of the parameters of the MHMM is extended from Equation~\ref{eq:L1} as
\begin{equation}\label{eq:submodels}
  \begin{aligned}
\log L &= \sum_{i=1}^N \log P(Y_i| \mathcal{M})\\
  &= \sum_{i=1}^N \log \left[ \sum_{k=1}^K P(\mathcal{M}^k) \sum_{\text{all } z} P\left(Y_i|z,\mathcal{M}^k\right) P\left(z|\mathcal{M}^k\right) \right]	 \\
	&= \sum_{i=1}^N \log \left[\sum_{k=1}^K w_k  \sum_{\text{all } z} \pi^k_{z_1} b^{k}_{z_1}(y_{i11})\cdots b^{k}_{z_1}(y_{i1C})
	  \prod_{t=2}^T \left[a^k_{z_{t-1}z_t} b^{k}_{z_t}(y_{it1})\cdots b^{k}_{z_t}(y_{itC}) \right] \right].
	  \end{aligned}
	\end{equation}

Compared to the usual hidden Markov model, there is an additional summation over the clusters in Equation~\ref{eq:submodels}, which seems to make the computations less straightforward than in the non-mixture case. Fortunately, by redefining MHMM as a special type HMM allows us to use standard HMM algorithms without major modifications. We combine the $K$ submodels into one large hidden Markov model consisting of $\sum_{k=1}^K S_k$ states, where the initial state vector contains elements of the form $w_k\pi^k$. Now the transition matrix is block diagonal
\begin{equation}
A =
 \begin{pmatrix}
  A^1   	& 0 			& \cdots & 0 			\\
  0 			& A^2 		& \cdots & 0 			\\
  \vdots  & \vdots  & \ddots & \vdots \\
  0 			& 0 			& \cdots & A^K
 \end{pmatrix},
\end{equation}
where the diagonal blocks $A^k, k=1,\ldots,K$, are square matrices containing the transition probabilities of one cluster. The off-diagonal blocks are zero matrices, so transitions between clusters are not allowed. Similarly, the emission matrices for each channel contain stacked emission matrices $B^k$.

\subsubsection{Covariates and cluster probabilities}\label{seq:covariates}

Covariates can be added to MHMM to explain cluster memberships as in latent class analysis. The prior cluster probabilities now depend on the subject's covariate values $\textbf{x}_i$ and are defined as multinomial distribution:
\begin{equation}
P(\mathcal{M}^k|\textbf{x}_i) = w_{ik}
= \frac{e^{\textbf{x}^{\top}_i\gamma_k}}{1 + \sum_{j=2}^K e^{\textbf{x}^{\top}_i\gamma_j}}.
\end{equation}
The first submodel is set as the reference by fixing $\gamma_1=(0,\ldots,0)^\top$.

As in MHMM without covariates, we can still use standard HMM algorithms with a slight modification; we now allow initial state probabilities $\pi$ to vary between subjects, i.e., for subject $i$ we have $\pi_i = (w_{i1} \pi^1, \ldots, w_{iK}\pi^K)^{\top}$. Of course, we also need to estimate the coefficients $\gamma$. For direct numerical maximization the modifications are straightforward. In the EM algorithm, regarding the M-step for $\gamma$, \pkg{seqHMM} uses iterative Newton's method with analytic gradients and Hessian which are straightforward to compute given all other model parameters. This Hessian can also be used for computing the conditional standard errors of coefficients. For unconditional standard errors, which take account of possible correlation between the estimates of $\gamma$ and other model parameters, the Hessian is computed using finite difference approximation of the Jacobian of the analytic gradients.

The posterior cluster probabilities $P(\mathcal{M}^k|Y_i, \textbf{x}_i)$ are obtained as
\begin{equation}
  \begin{aligned}
P(\mathcal{M}^k|Y_i, \textbf{x}_i) &= \frac{P(Y_i | \mathcal{M}^k, \textbf{x}_i) P(\mathcal{M}^k|\textbf{x}_i) }{P(Y_i| \textbf{x}_i)}\\
&= \frac{P(Y_i | \mathcal{M}^k, \textbf{x}_i) P(\mathcal{M}^k|\textbf{x}_i) }{\sum_{j=1}^K P(Y_i |\mathcal{M}^j, \textbf{x}_i) P(\mathcal{M}^j|\textbf{x}_i)} = \frac{L_k^i}{L^i},
  \end{aligned}
\end{equation}
where $L^i$ is the likelihood of the complete MHMM for subject $i$, and $L_k^i$ is the likelihood of cluster $k$ for subject $i$. These are straightforwardly computed from forward probabilities. Posterior cluster probabilities are used e.g., for computing classification tables.

\subsection{Important special cases}

The hidden Markov model is not the only important special case of the mixture hidden Markov model. Here we cover some of the most important special cases that are included in the \pkg{seqHMM} package.

\subsubsection{Markov model}

The Markov model (MM) is a special case of the HMM, where there is no hidden structure. It can be regarded as an HMM where the hidden states correspond to the observed states perfectly. Now the number of hidden states matches the number of the observed states. The emission probability $P(y_{it}) = 1$ if $z_t = y_{it}$ and $0$ otherwise, i.e., the emission matrices are identity matrices. Note that for building Markov models the data must be in a single-channel format.

\subsubsection{Mixture Markov model}

Like MM, the mixture Markov model (MMM) is a special case of the MHMM, where there is no hidden structure. The likelihood of the model is now of the form
\begin{equation}
  \begin{aligned}
\log L &= \sum_{i = 1}^N \log P(\mathbf{y}_i | \mathbf{x}_i, \mathcal{M}^k) = \sum_{i = 1}^N \log \sum_{k=1}^K P(\mathcal{M}^k|\mathbf{x}_i) P(\mathbf{y}_i | \mathbf{x}_i, \mathcal{M}^k)\\
&= \sum_{i = 1}^N \log \sum_{k=1}^K P(\mathcal{M}^k|\mathbf{x}_i) P(y_{i1}|\mathbf{x}_i, \mathcal{M}^k) \prod_{t = 2}^T P(y_{it} | y_{i(t-1)}, \mathbf{x}_i, \mathcal{M}^k).
  \end{aligned}
\end{equation}
Again, the data must be in a single-channel format.

\subsubsection{Latent class model}

Latent class models (LCM) are another class of models that are often used for longitudinal research. Such models have been called, e.g., (latent) growth models, latent trajectory models, or longitudinal latent class models \citep{Vermunt2008, Collins1992}. These models assume that dependencies between observations can be captured by a latent class, i.e., a time-constant variable which we call cluster in this paper.

The \pkg{seqHMM} includes a function for fitting an LCM as a special case of MHMM where there is only one hidden state for each cluster. The transition matrix of each cluster is now reduced to a scalar 1 and the likelihood is of the form
\begin{equation}
  \begin{aligned}
\log L &= \sum_{i = 1}^N \log P(Y_i | \mathbf{x}_i, \mathcal{M}^k)
= \sum_{i = 1}^N \log \sum_{k=1}^K P(\mathcal{M}^k|\mathbf{x}_i) P(Y_i | \mathbf{x}_i, \mathcal{M}^k)\\
&= \sum_{i = 1}^N \log \sum_{k=1}^K P(\mathcal{M}^k|\mathbf{x}_i) \prod_{t = 1}^T P(\textbf{y}_{it} | \mathbf{x}_i, \mathcal{M}^k).
  \end{aligned}
\end{equation}

For LCMs, the data can consist of multiple channels, i.e., the data for each subject consists of multiple parallel sequences. It is also possible to use \pkg{seqHMM} for estimating LCMs for non-longitudinal data with only one time point, e.g., to study multiple questions in a survey.

\section{Package features}

The purpose of the \pkg{seqHMM} package is to offer tools for the whole HMM analysis process from sequence data manipulation and description to model building, evaluation, and visualization. Naturally, \pkg{seqHMM} builds on other packages, especially the \pkg{TraMineR} package designed for sequence analysis. For constructing, summarizing, and visualizing sequence data, \pkg{TraMineR} provides many useful features. First of all, we use the \pkg{TraMineR}'s \code{stslist} class as the sequence data structure of \pkg{seqHMM}. These state sequence objects have attributes such as color palette and alphabet, and they have specific methods for plotting, summarizing, and printing. Many other \pkg{TraMineR}'s features for plotting or data manipulation are also used in \pkg{seqHMM}.

On the other hand, \pkg{seqHMM} extends the functionalities of \pkg{TraMineR}, e.g., by providing easy-to-use plotting functions for multichannel data and a simple function for converting such data into a single-channel representation.

Other significant packages used by \pkg{seqHMM} include the \pkg{igraph} package \citep{Csardi2006}, which is used for drawing graphs of HMMs, and the \pkg{nloptr} package \citep{nloptr, nlopt}, which is used in direct numerical optimization of model parameters. The computationally intensive parts of the package are written in \proglang{C++} with the help of the \pkg{Rcpp} \citep{Eddelbuettel2011, Eddelbuettel2013} and \pkg{RcppArmadillo} \citep{Eddelbuettel2014} packages. In addition to using \proglang{C++} for major algorithms, \pkg{seqHMM} also supports parallel computation via the OpenMP interface \citep{Dagum1998} by dividing computations for subjects between threads.

\begin{table}
\begin{tabular}[c]{p{0.4\textwidth} L{0.55\textwidth}}
\hline
Usage               & Functions/methods \\
\hline
Model construction & \code{build\_hmm}, \code{build\_mhmm}, \code{build\_mm},
\code{build\_mmm}, \code{build\_lcm},
\code{simulate\_initial\_probs}, \code{simulate\_transition\_probs}, \code{simulate\_emission\_probs} \\
\hline
Model estimation      & \code{fit\_model} \\
\hline
Model visualization     & \code{plot}, \code{ssplot}, \code{mssplot} \\
\hline
Model inference     & \code{logLik}, \code{BIC}, \code{summary} \\
\hline
State inference     & \code{hidden_paths}, \code{posterior_probs}, \code{forward_backward} \\
\hline
Data visualization       & \code{ssplot}, \code{ssp} + \code{plot}, \code{ssp} + \code{gridplot}\\
\hline
Data and model manipulation & \code{mc\_to\_sc}, \code{mc\_to\_sc\_data}, \code{trim\_model}, \code{separate\_mhmm} \\
\hline
Data simulation   & \code{simulate\_hmm}, \code{simulate\_mhmm} \\
\hline

\end{tabular}
\caption{Functions and methods in the \pkg{seqHMM} package.}
\label{table:functions}
\end{table}

Table \ref{table:functions} shows the functions and methods available in the \pkg{seqHMM} package. The package includes functions for estimating and evaluating HMMs and MHMMs as well as visualizing data and models. There are some functions for manipulating data and models, and for simulating model parameters or sequence data given a model. In the next sections we discuss the usage of these functions more thoroughly.

As the straightforward implementation of the forward--backward algorithm poses a great risk of under- and overflow, typically forward probabilities are scaled so that there should be no underflow. \pkg{seqHMM} uses the scaling as in \citet{Rabiner1989}, which is typically sufficient for numerical stability. In case of MHMM though, we have sometimes observed numerical issues in the forward algorithm even with proper scaling. Fortunately this usually means that the backward algorithm fails completely, giving a clear signal that something is wrong. This is especially true in the case of global optimization algorithms which can search unfeasible areas of the parameter space, or when using bad initial values often with large number of zero-constraints. Thus, \pkg{seqHMM} also supports computation on the logarithmic scale in most of the algorithms, which further reduces the numerical unstabilities. On the other hand, as there is a need to back-transform to the natural scale during the algorithms, the log-space approach is somewhat slower than the scaling approach. Therefore, the default option is to use the scaling approach, which can be changed to the log-space approach by setting the \code{log\_space} argument to \code{TRUE} in, e.g., \code{fit\_model}.

\subsection{Building and fitting models}

A model is first constructed using an appropriate build function. As Table \ref{table:functions} illustrates, several such functions are available: \code{build\_hmm} for hidden Markov models, \code{build\_mhmm} for mixture hidden Markov models, \code{build\_mm} for Markov models, \code{build\_mmm} for mixture Markov models, and \code{build\_lcm} for latent class models.

The user may give their own starting values for model parameters, which is typically advisable for improved efficiency, or use random starting values. Build functions check that the data and parameter matrices (when given) are of the right form and create an object of class \code{hmm} (for HMMs and MMs) or \code{mhmm} (for MHMMs, MMMs, and LCMs). For ordinary Markov models, the \code{build\_mm} function automatically estimates the initial probabilities and the transition matrix based on the observations. For this type of model, starting values or further estimation are not needed. For mixture models, covariates can be omitted or added with the usual \code{formula} argument using symbolic formulas familiar from, e.g., the \code{lm} function. Even though missing observations are allowed in sequence data, covariates must be completely observed.

After a model is constructed, model parameters may be estimated with the \code{fit\_model} function. MMs, MMMs, and LCMs are handled internally as their more general counterparts, except in the case of \code{print} methods, where some redundant parts of the model are not printed.

In all models, initial zero probabilities are regarded as structural zeroes and only positive probabilities are estimated. Thus it is easy to construct, e.g., a left-to-right model by defining the transition probability matrix as an upper triangular matrix.

The \code{fit\_model} function provides three estimation steps: 1) EM algorithm, 2) global DNM, and 3) local DNM. The user can call for one method or any combination of these steps, but should note that they are performed in the above-mentioned order. At the first step, starting values are based on the model object given to \code{fit\_model}. Results from a former step are then used as starting values in the latter. Exceptions to this rule include some global optimization algorithms, which do not use initial values (because of this, performing just the local DNM step can lead to a better solution than global DNM with a small number of iterations). 

We have used our own implementation of the EM algorithm for MHMMs whereas the DNM steps (2 and 3) rely on the optimization routines provided by the \pkg{nloptr} package. The EM algorithm and computation of gradients were written in \proglang{C++} with an option for parallel computation between subjects. The user can choose the number of parallel threads (typically, the number of cores) with the \code{threads} argument.

In order to reduce the risk of being trapped in a poor local optimum, a large number of initial values should be tested. The \pkg{seqHMM} package strives to automatize this. One option is to run the EM algorithm multiple times with more or less random starting values for transition or emission probabilities or both. These are called for in the \code{control\_em} argument. Although not done by default, this method seems to perform very well as the EM algorithm is relatively fast compared to DNM.

Another option is to use a global DNM approach such as the multilevel single-linkage method (MLSL) \citep{Kan1987I, Kan1987II}. It draws multiple random starting values and performs local optimization from each starting point. The LDS modification uses low-discrepancy sequences instead of random numbers as starting points and should improve the convergence rate \citep{Kucherenko2005}.

By default, the \code{fit\_model} function uses the EM algorithm with a maximum of 1000 iterations and skips the local and global DNM steps. For the local step, the L-BFGS algorithm \citep{Nocedal1980, Liu1989} is used by default. Setting \code{global\_step = TRUE}, the function performs MSLS-LDS with the L-BFGS as the local optimizer. In order to reduce the computation time spent on non-global optima, the convergence tolerance of the local optimizer is set relatively large, so again local optimization should be performed at the final step.

Unfortunately, there is no universally best optimization method. For unconstrained problems, the computation time for a single EM or DNM rapidly increases as the model size increases and at the same time the risk of getting trapped in a local optimum or a saddle point also increases. As \pkg{seqHMM} provides functions for analytic gradients, the optimization routines of \pkg{nloptr} which make use of this information are likely preferable. In practice we have had most success with randomized EM, but it is advisable to try a couple of different settings; e.g., randomized EM, EM followed by global DNM, and only global DNM, perhaps with different optimization routines. Documentation of the \code{fit_model} function gives examples of different optimization strategies and how they can lead to different solutions.

For examples on model estimation and starting values, see a supplementary vignette on CRAN \citep{Helske2017b}.

\subsubsection{State and model inference}

In \pkg{seqHMM}, forward and backward probabilities are computed using the \code{forward\_backward} function, either on the logarithmic scale or in the form of scaled probabilities, depending on the argument \code{log\_space}. Posterior probabilities are obtained from the \code{posterior_probs} function. In \pkg{seqHMM}, the most probable paths are computed with the \code{hidden\_paths} function. For details of the Viterbi and the forward--backward algorithm, see e.g., \citet{Rabiner1989}.

The \pkg{seqHMM} package provides the \code{logLik} method for computing the log-likelihood of a model. The method returns an object of class \code{logLik} which is compatible with the generic information criterion functions \code{AIC} and \code{BIC} of \proglang{R}. When constructing the \code{hmm} and \code{mhmm} objects via model building functions, the number of observations and the number of parameters of the model are stored as attributes \code{nobs} and \code{df} which are extracted by the \code{logLik} method for the computation of information criteria. The number of model parameters defined from the initial model by taking account of the parameter redundancy constraints (stemming from sum-to-one constraints of transition, emission, and initial state probabilities) and by defining all zero probabilities as structural, fixed values.

The \code{summary} method automatically computes some features for the MHMM, MMM, and the latent class model, e.g., standard errors for covariates and prior and posterior cluster probabilities for subjects. A \code{print} method for this summary shows an output of the summaries: estimates and standard errors for covariates, log-likelihood and BIC, and information on most probable clusters and prior probabilities.

\subsection{Visualizing sequence data}
\label{sec:vizualizing}

Good graphical presentations of data and models are useful during the whole analysis process from the first glimpse into the data to the model fitting and presentation of results. The \pkg{TraMineR} package provides nice plotting options and summaries for simple sequence data, but at the moment there is no easy way of plotting multichannel data. We propose to use a so-called \emph{stacked sequence plot} (ssp), where the channels are plotted on top of each other so that the same row in each figure matches the same subject. Figure \ref{fig:graphicalillustrations2} illustrates an example of a stacked sequence plot with the ten first sequences of the \code{biofam} data set. The code for creating the figure is shown in Section \hyperlink{plottingI}{4.1}.

\begin{knitrout}
\definecolor{shadecolor}{rgb}{0.969, 0.969, 0.969}\color{fgcolor}\begin{figure}

{\centering \includegraphics[width=\maxwidth]{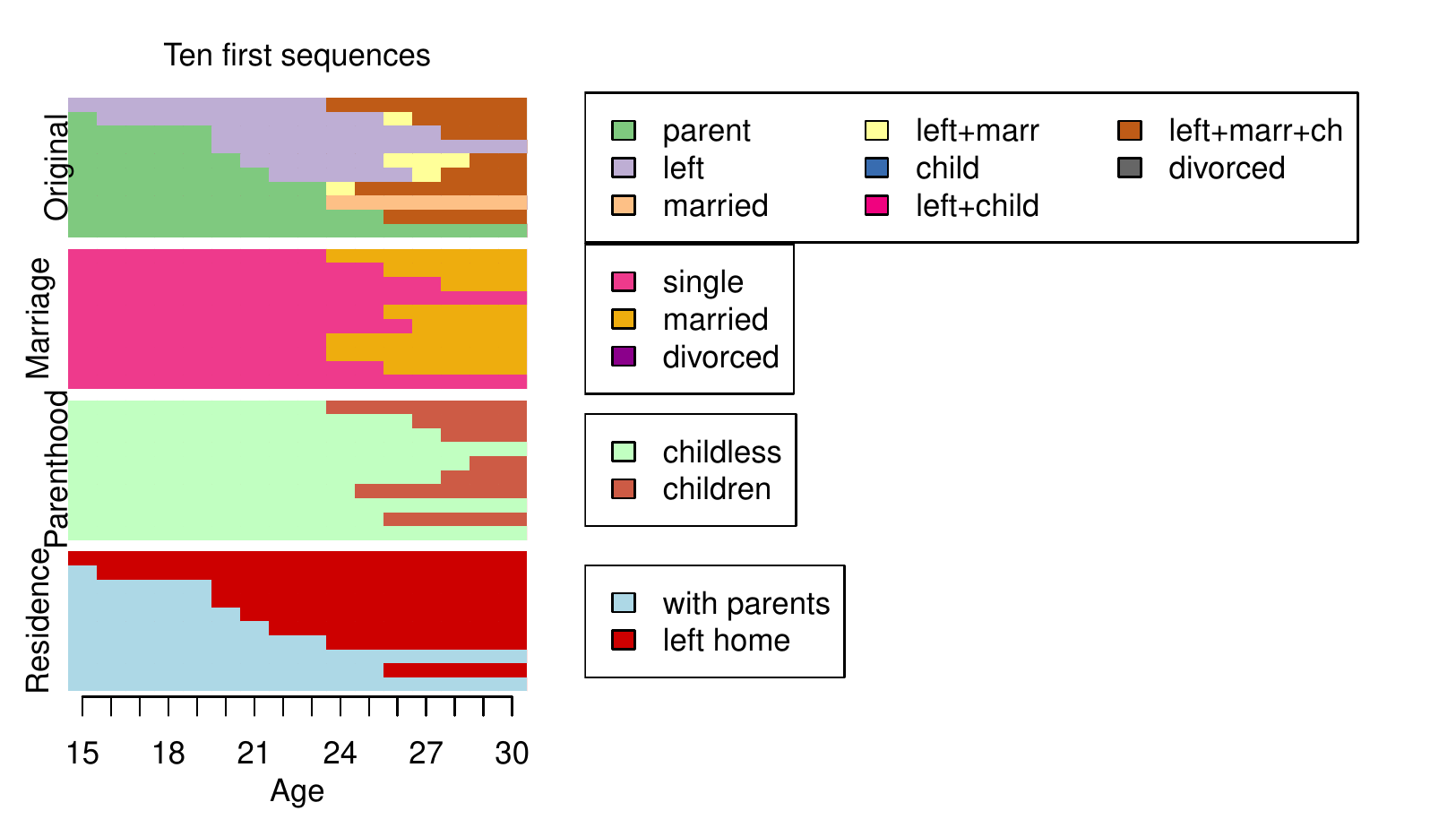} 

}

\caption[Stacked sequence plot of the first ten individuals in the \code{biofam} data plotted with the \code{ssplot} function]{Stacked sequence plot of the first ten individuals in the \code{biofam} data plotted with the \code{ssplot} function. The top plot shows the original sequences, and the three bottom plots show the sequences in the separate channels for the same individuals. The sequences are in the same order in each plot, i.e., the same row always matches the same individual.}\label{fig:graphicalillustrations2}
\end{figure}

\end{knitrout}

The \code{ssplot} function is the simplest way of plotting multichannel sequence data in \pkg{seqHMM}. It can be used to illustrate state distributions or sequence index plots. The former is the default option, since index plots can take a lot of time and memory if data are large. Figure \ref{fig:plottingsequences} illustrates a default plot which the user can modify in many ways (see the code in Section \hyperlink{plottingd}{4.1}). More examples are shown in the documentation pages of the \code{ssplot} function.

\begin{knitrout}
\definecolor{shadecolor}{rgb}{0.969, 0.969, 0.969}\color{fgcolor}\begin{figure}

{\centering \includegraphics[width=\maxwidth]{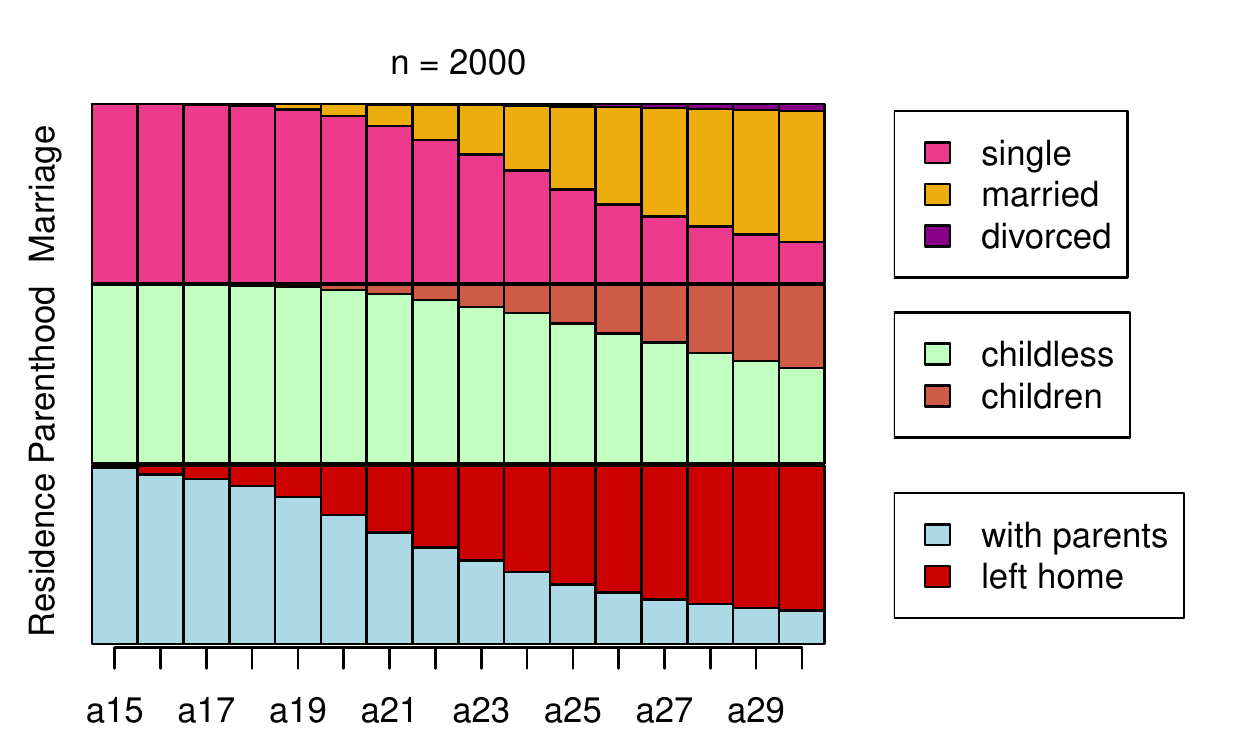} 

}

\caption{Stacked sequence plot of annual state distributions in the three-channel \code{biofam} data. This is the default output of the \code{ssplot} function. The labels for the channels are taken from the named list of state sequence objects, and the labels for the x axis ticks are taken from the column names of the first object.}\label{fig:plottingsequences}
\end{figure}

\end{knitrout}

Another option is to define function arguments with the \code{ssp} function and then use previously saved arguments for plotting with a simple \code{plot} method. It is also possible to combine several \code{ssp} figures into one plot with the \code{gridplot} function. Figure \ref{fig:gridplot1} illustrates an example of such a plot showing sequence index plots for women and men (see the code in Section \hyperlink{plottinggrid}{4.1}). Sequences are ordered in a more meaningful order using multidimensional scaling scores of observations (computed from sequence dissimilarities). After defining the plot for one group, a similar plot for other groups is easily defined using the \code{update} function.

\begin{knitrout}
\definecolor{shadecolor}{rgb}{0.969, 0.969, 0.969}\color{fgcolor}\begin{figure}

{\centering \includegraphics[width=\maxwidth]{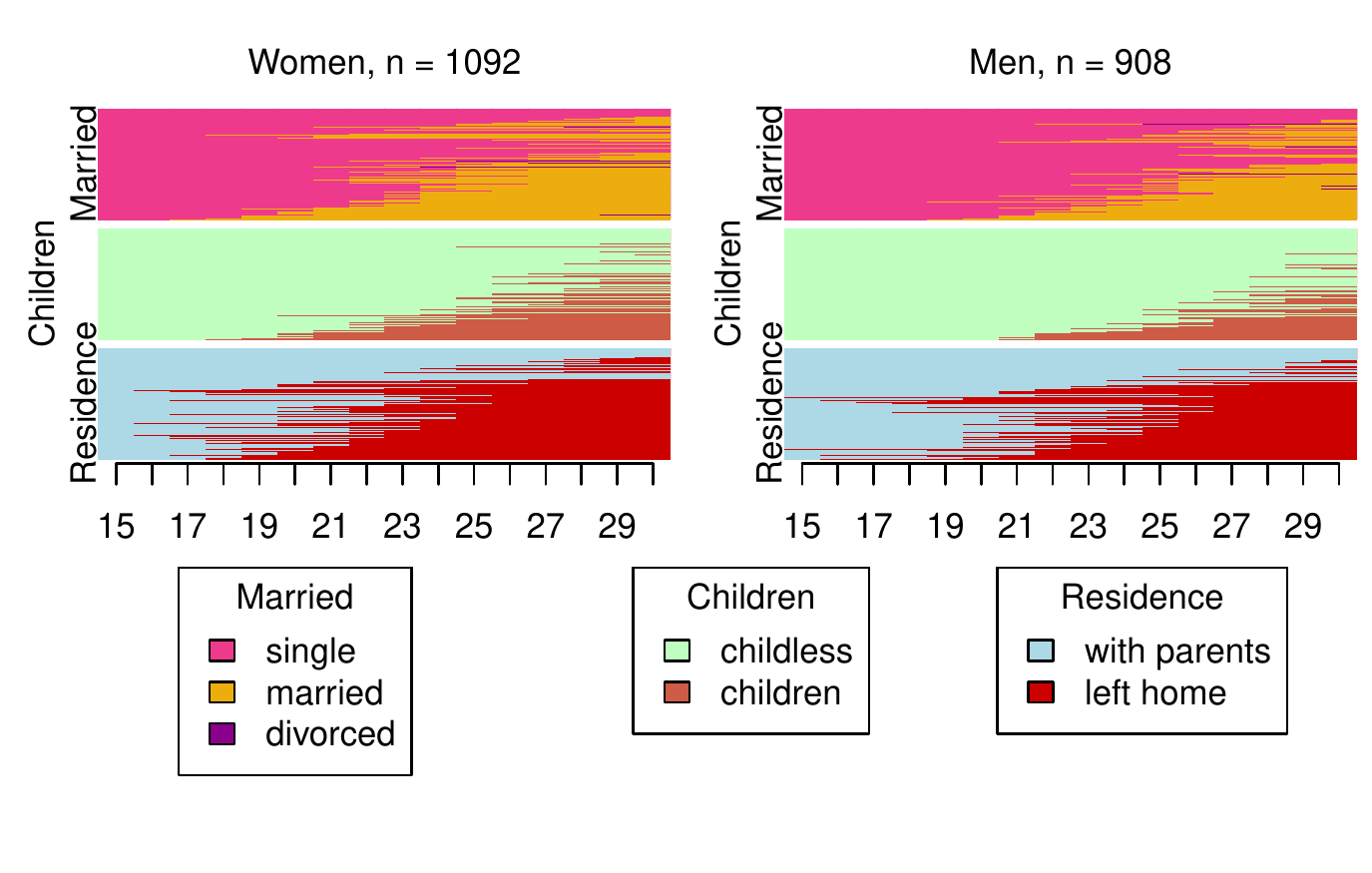} 

}

\caption{Showing state distribution plots for women and men in the \code{biofam} data. Two figures were defined with the \code{ssp} function and then combined into one figure with the \code{gridplot} function.}\label{fig:gridplot1}
\end{figure}

\end{knitrout}

The \code{gridplot} function is useful for showing different features for the same subjects or the same features for different groups. The user has a lot of control over the layout, e.g., dimensions of the grid, widths and heights of the cells, and positions of the legends.

We also provide a function \code{mc\_to\_sc\_data} for the easy conversion of multichannel sequence data into a single channel representation. Plotting combined data is often useful in addition to (or instead of) showing separate channels.

\subsection{Visualizing hidden Markov models}
\label{seq:vizHMM}

For the easy visualization of the model structure and parameters, we propose plotting HMMs as directed graphs. Such graphs are easily called with the \code{plot} method, with an object of class \code{hmm} as an argument. Figure \ref{fig:plottingHMM} illustrates a five-state HMM. The code for producing the plot is shown in Section \hyperlink{code_plottingHMM}{4.4}.

\begin{knitrout}
\definecolor{shadecolor}{rgb}{0.969, 0.969, 0.969}\color{fgcolor}\begin{figure}

{\centering \includegraphics[width=\linewidth]{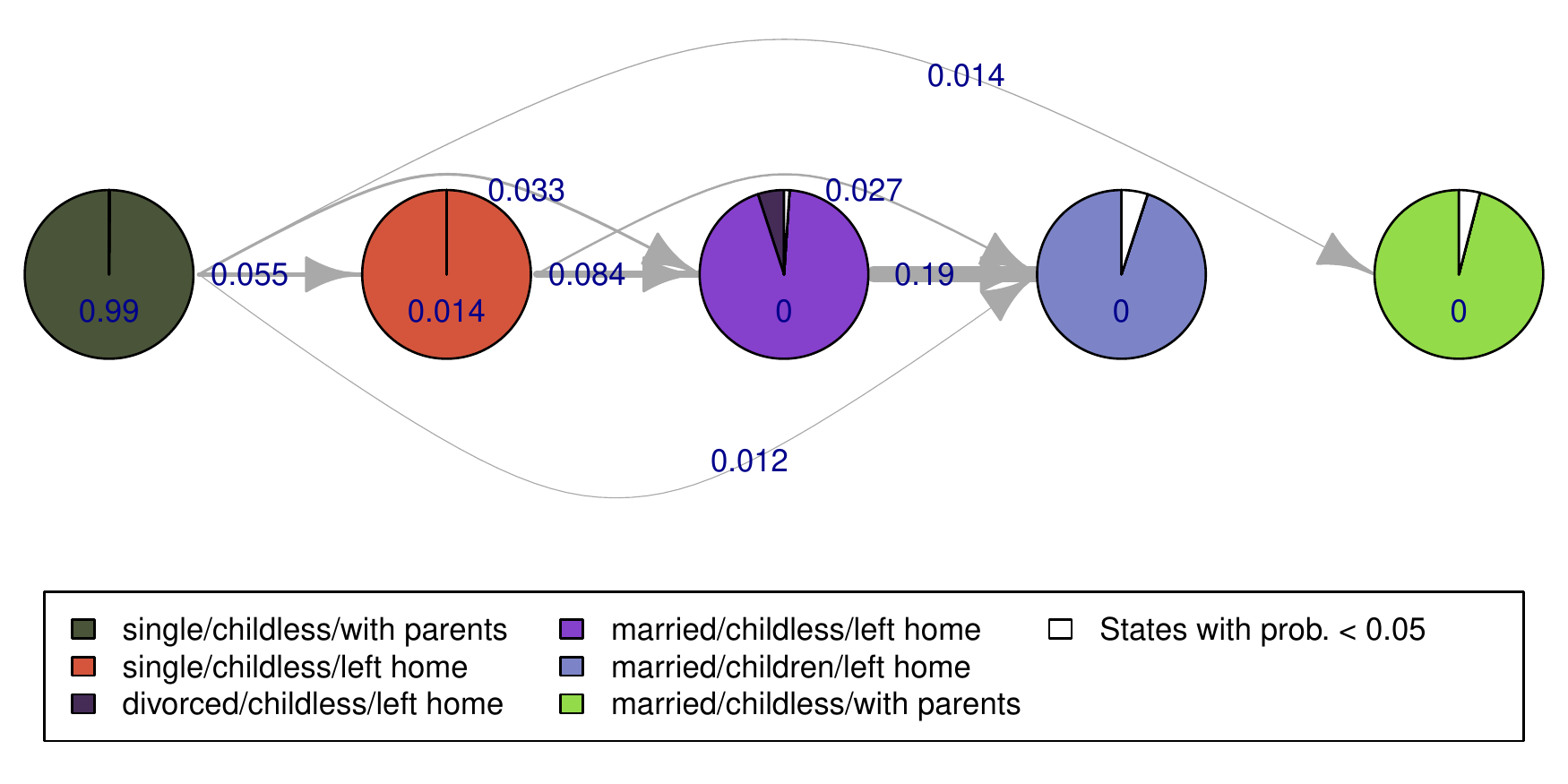} 

}

\caption[Illustrating a hidden Markov model as a directed graph]{Illustrating a hidden Markov model as a directed graph. Pies represent five hidden states, with slices showing emission probabilities of combinations of observed states. States with emission probability less than 0.05 are combined into one slice. Edges show the transtion probabilities. Initial probabilities of hidden states are given below the pies.}\label{fig:plottingHMM}
\end{figure}

\end{knitrout}

Hidden states are presented with pie charts as vertices (or nodes), and transition probabilities are shown as edges (arrows, arcs). By default, the higher the transition probability, the thicker the stroke of the edge. Emitted observed states are shown as slices in the pies. For gaining a simpler view, observations with small emission probabilities (less than 0.05 by default) can be combined into one category. Initial state probabilities are given below or next to the respective vertices. In the case of multichannel sequences, the data and the model are converted into a single-channel representation with the \code{mc\_to\_sc} function.

A simple default plot is easy to call, but the user has a lot of control over the layout. Figure \ref{fig:graphicalillustrations5} illustrates another possible visualization of the same model. The code is shown in Section \hyperlink{code_graphicalillustrations5}{4.4}.

\begin{knitrout}
\definecolor{shadecolor}{rgb}{0.969, 0.969, 0.969}\color{fgcolor}\begin{figure}

{\centering \includegraphics[width=\linewidth]{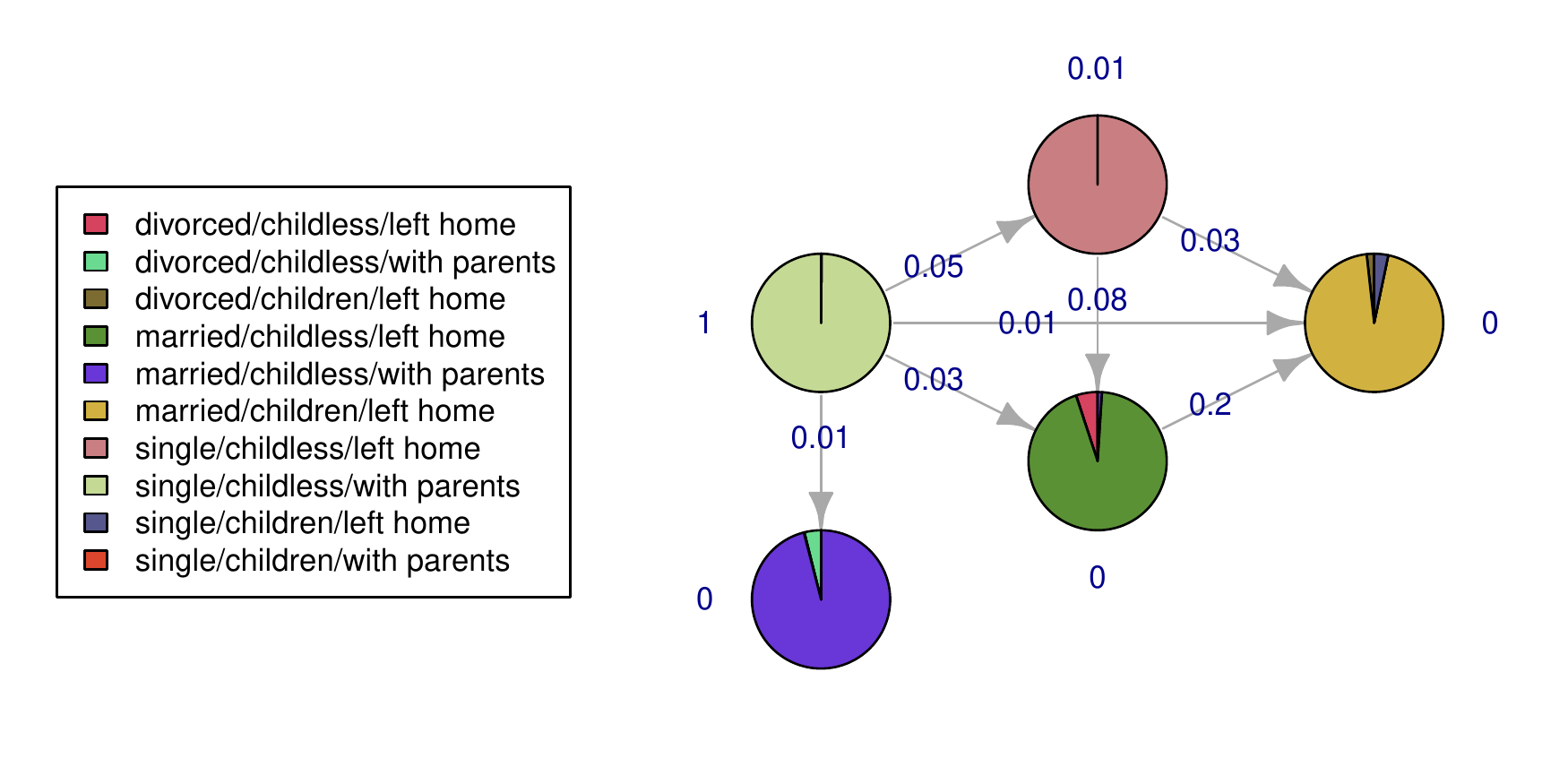} 

}

\caption[Another version of the hidden Markov model of Figure 4 with a different layout and modified labels, legends, and colors]{Another version of the hidden Markov model of Figure 4 with a different layout and modified labels, legends, and colors. All observed states are shown.}\label{fig:graphicalillustrations5}
\end{figure}

\end{knitrout}

The \code{ssplot} function (see Section \ref{sec:vizualizing}) also accepts an object of class \code{hmm}. The user can easily choose to plot observations, most probable paths of hidden states, or both. The function automatically computes hidden paths if the user does not provide them.

Figure \ref{fig:ssplotHMM} shows observed sequences with the most probable paths of hidden states given the model. Sequences are sorted according to multidimensional scaling scores computed from hidden paths. The code for creating the plot is shown in Section \hyperlink{ssplotHMM}{4.4}.

\begin{knitrout}
\definecolor{shadecolor}{rgb}{0.969, 0.969, 0.969}\color{fgcolor}\begin{figure}

{\centering \includegraphics[width=\maxwidth]{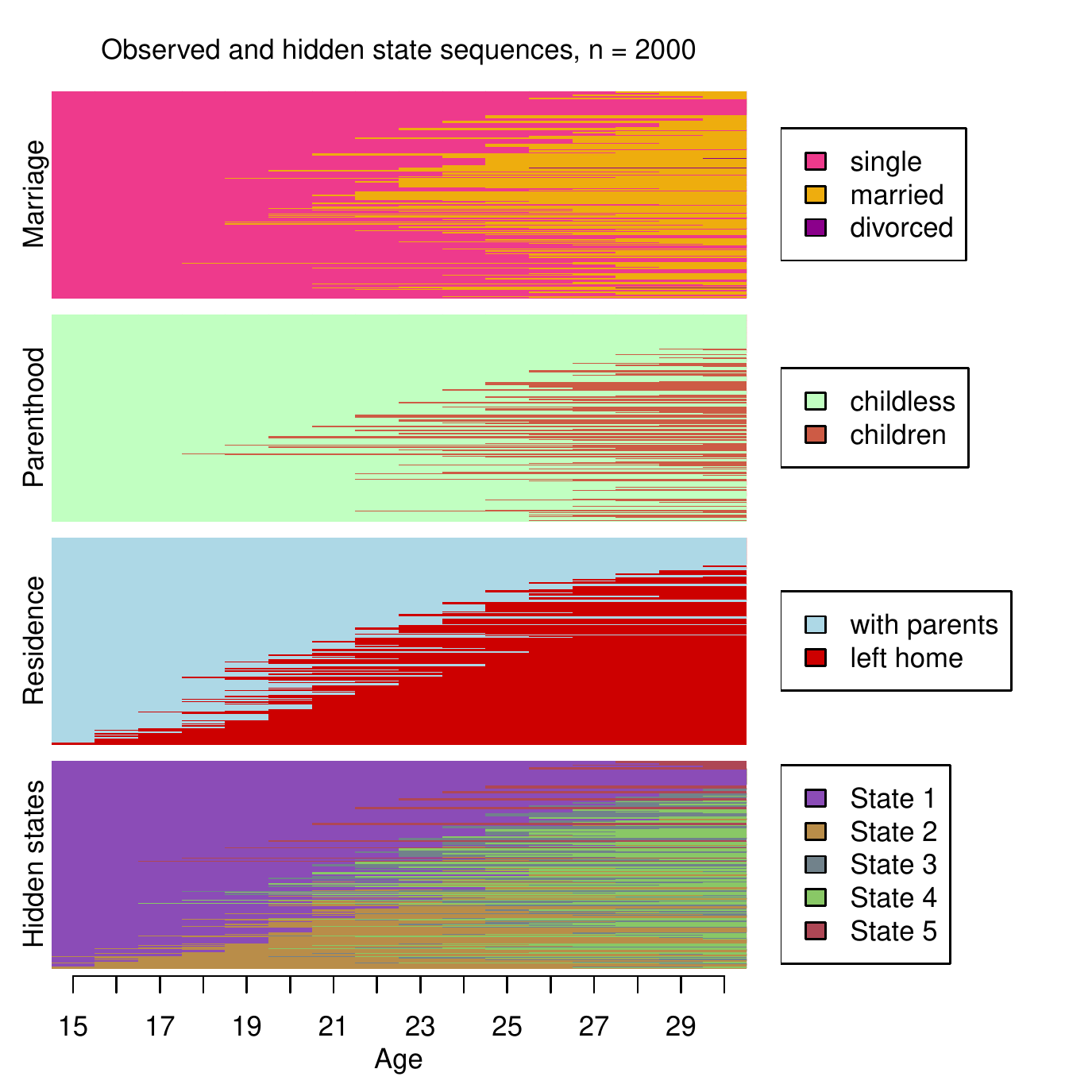} 

}

\caption[Using the \code{ssplot} function for an \code{hmm} object makes it easy to plot the observed sequences together with the most probable paths of hidden states given the model]{Using the \code{ssplot} function for an \code{hmm} object makes it easy to plot the observed sequences together with the most probable paths of hidden states given the model.}\label{fig:ssplotHMM}
\end{figure}

\end{knitrout}

The \code{plot} method works for \code{mhmm} objects as well. The user can choose between an interactive mode, where the model for each (chosen) cluster is plotted separately, and a combined plot with all models in one plot. The equivalent to the \code{ssplot} function for MHMMs is \code{mssplot}. It plots stacked sequence plots separately for each cluster. If the user asks to plot more than one cluster, the function is interactive by default.

\section{Examples with life course data}
\label{sec:code}

In this section we show examples of using the \pkg{seqHMM} package. We start by constructing and visualizing sequence data, then show how HMMs are built and fitted for single-channel and multichannel data, then move on to clustering with MHMMs, and finally illustrate how to plot HMMs.

Throughout the examples we use the same \code{biofam} data described in Section \ref{sec:seqdata}. We use both the original single-channel data and a three-channel modification named \code{biofam3c}, which is included in the \pkg{seqHMM} package. For more information on the conversion, see the documentation of the \code{biofam3c} data.

\subsection{Sequence data}
\label{sec:exseqdata}

Before getting to the estimation, it is good to get to know the data. We start by loading the original \code{biofam} data as well as the three-channel version of the same data, \code{biofam3c}. We convert the data into the \code{stslist} form with the \code{seqdef} function. We set the starting age at 15 and set the order of the states with the \code{alphabet} argument (for plotting). Colors of the states can be modified and stored as an attribute in the \code{stslist} object -- this way the user only needs to define them once.

\begin{knitrout}
\definecolor{shadecolor}{rgb}{0.969, 0.969, 0.969}\color{fgcolor}\begin{kframe}
\begin{alltt}
\hlkwd{library}\hlstd{(}\hlstr{"seqHMM"}\hlstd{)}

\hlkwd{data}\hlstd{(}\hlstr{"biofam"}\hlstd{,} \hlkwc{package} \hlstd{=} \hlstr{"TraMineR"}\hlstd{)}
\hlstd{biofam_seq} \hlkwb{<-} \hlkwd{seqdef}\hlstd{(biofam[,} \hlnum{10}\hlopt{:}\hlnum{25}\hlstd{],} \hlkwc{start} \hlstd{=} \hlnum{15}\hlstd{,} \hlkwc{labels} \hlstd{=} \hlkwd{c}\hlstd{(}\hlstr{"parent"}\hlstd{,}
  \hlstr{"left"}\hlstd{,} \hlstr{"married"}\hlstd{,} \hlstr{"left+marr"}\hlstd{,} \hlstr{"child"}\hlstd{,} \hlstr{"left+child"}\hlstd{,} \hlstr{"left+marr+ch"}\hlstd{,}
  \hlstr{"divorced"}\hlstd{))}

\hlkwd{data}\hlstd{(}\hlstr{"biofam3c"}\hlstd{)}
\hlstd{marr_seq} \hlkwb{<-} \hlkwd{seqdef}\hlstd{(biofam3c}\hlopt{$}\hlstd{married,} \hlkwc{start} \hlstd{=} \hlnum{15}\hlstd{,} \hlkwc{alphabet} \hlstd{=} \hlkwd{c}\hlstd{(}\hlstr{"single"}\hlstd{,}
  \hlstr{"married"}\hlstd{,} \hlstr{"divorced"}\hlstd{))}
\hlstd{child_seq} \hlkwb{<-} \hlkwd{seqdef}\hlstd{(biofam3c}\hlopt{$}\hlstd{children,} \hlkwc{start} \hlstd{=} \hlnum{15}\hlstd{,}
  \hlkwc{alphabet} \hlstd{=} \hlkwd{c}\hlstd{(}\hlstr{"childless"}\hlstd{,} \hlstr{"children"}\hlstd{))}
\hlstd{left_seq} \hlkwb{<-} \hlkwd{seqdef}\hlstd{(biofam3c}\hlopt{$}\hlstd{left,} \hlkwc{start} \hlstd{=} \hlnum{15}\hlstd{,} \hlkwc{alphabet} \hlstd{=} \hlkwd{c}\hlstd{(}\hlstr{"with parents"}\hlstd{,}
  \hlstr{"left home"}\hlstd{))}

\hlkwd{attr}\hlstd{(marr_seq,} \hlstr{"cpal"}\hlstd{)} \hlkwb{<-} \hlkwd{c}\hlstd{(}\hlstr{"violetred2"}\hlstd{,} \hlstr{"darkgoldenrod2"}\hlstd{,} \hlstr{"darkmagenta"}\hlstd{)}
\hlkwd{attr}\hlstd{(child_seq,} \hlstr{"cpal"}\hlstd{)} \hlkwb{<-} \hlkwd{c}\hlstd{(}\hlstr{"darkseagreen1"}\hlstd{,} \hlstr{"coral3"}\hlstd{)}
\hlkwd{attr}\hlstd{(left_seq,} \hlstr{"cpal"}\hlstd{)} \hlkwb{<-} \hlkwd{c}\hlstd{(}\hlstr{"lightblue"}\hlstd{,} \hlstr{"red3"}\hlstd{)}
\end{alltt}
\end{kframe}
\end{knitrout}

Here we show codes for creating Figures \ref{fig:plottingsequences}, \ref{fig:graphicalillustrations2}, and \ref{fig:gridplot1}. Such plots give a good glimpse into multichannel data.

\hypertarget{plottingd}{}\subsubsection[Figure 2: Plotting state distributions]{Figure \ref{fig:plottingsequences}: Plotting state distributions}

We start by showing how to call the simple default plot of Figure \ref{fig:plottingsequences} in Section \ref{seq:vizHMM}. By default the function plots state distributions (\code{type = "d"}). Multichannel data are given as a list where each component is an \code{stslist} corresponding to one channel. If names are given, those will be used as labels in plotting.

\begin{knitrout}
\definecolor{shadecolor}{rgb}{0.969, 0.969, 0.969}\color{fgcolor}\begin{kframe}
\begin{alltt}
\hlkwd{ssplot}\hlstd{(}\hlkwd{list}\hlstd{(}\hlstr{"Marriage"} \hlstd{= marr_seq,} \hlstr{"Parenthood"} \hlstd{= child_seq,}
  \hlstr{"Residence"} \hlstd{= left_seq))}
\end{alltt}
\end{kframe}
\end{knitrout}

\hypertarget{plottingI}{}\subsubsection[Figure 1: Plotting sequences]{Figure \ref{fig:graphicalillustrations2}: Plotting sequences}

Figure \ref{fig:graphicalillustrations2} with the whole sequences requires modifying more arguments. We call for sequence index plots (\code{type = "I"}) and sort sequences according to the first channel (the original sequences), starting from the beginning. We give labels to y and x axes and modify the positions of y labels. We give a title to the plot but omit the number of subjects, which by default is printed. We set the proportion of the plot given to legends and the number of columns in each legend.

\begin{knitrout}
\definecolor{shadecolor}{rgb}{0.969, 0.969, 0.969}\color{fgcolor}\begin{kframe}
\begin{alltt}
\hlkwd{ssplot}\hlstd{(}\hlkwd{list}\hlstd{(biofam_seq[}\hlnum{1}\hlopt{:}\hlnum{10}\hlstd{,], marr_seq[}\hlnum{1}\hlopt{:}\hlnum{10}\hlstd{,], child_seq[}\hlnum{1}\hlopt{:}\hlnum{10}\hlstd{,],}
  \hlstd{left_seq[}\hlnum{1}\hlopt{:}\hlnum{10}\hlstd{,]),}
  \hlkwc{sortv} \hlstd{=} \hlstr{"from.start"}\hlstd{,} \hlkwc{sort.channel} \hlstd{=} \hlnum{1}\hlstd{,} \hlkwc{type} \hlstd{=} \hlstr{"I"}\hlstd{,}
  \hlkwc{ylab} \hlstd{=} \hlkwd{c}\hlstd{(}\hlstr{"Original"}\hlstd{,} \hlstr{"Marriage"}\hlstd{,} \hlstr{"Parenthood"}\hlstd{,} \hlstr{"Residence"}\hlstd{),}
  \hlkwc{xtlab} \hlstd{=} \hlnum{15}\hlopt{:}\hlnum{30}\hlstd{,} \hlkwc{xlab} \hlstd{=} \hlstr{"Age"}\hlstd{,} \hlkwc{title} \hlstd{=} \hlstr{"Ten first sequences"}\hlstd{,}
  \hlkwc{title.n} \hlstd{=} \hlnum{FALSE}\hlstd{,} \hlkwc{legend.prop} \hlstd{=} \hlnum{0.63}\hlstd{,} \hlkwc{ylab.pos} \hlstd{=} \hlkwd{c}\hlstd{(}\hlnum{1}\hlstd{,} \hlnum{1.5}\hlstd{),}
  \hlkwc{ncol.legend} \hlstd{=} \hlkwd{c}\hlstd{(}\hlnum{3}\hlstd{,} \hlnum{1}\hlstd{,} \hlnum{1}\hlstd{,} \hlnum{1}\hlstd{))}
\end{alltt}
\end{kframe}
\end{knitrout}

\hypertarget{plottinggrid}{}\subsubsection[Figure 3: Plotting sequence data in a grid]{Figure \ref{fig:gridplot1}: Plotting sequence data in a grid}

For using the \code{gridplot} function, we first need to specify the \code{ssp} objects of the separate plots. Here we start by defining the first plot for women with the \code{ssp} function. It stores the features of the plot, but does not draw anything. We want to sort sequences according to multidimensional scaling scores. These are computed from optimal matching dissimilarities for observed sequences. Any dissimilarity method available in \code{TraMineR} can be used instead of the default (see the documentation of the \code{seqdef} function for more information). We want to use the same legends for the both plots, so we remove legends from the \code{ssp} objects.

Since we are going to plot to two similar figures, one for women and one for men, we can pass the first \code{ssp} object to the \code{update} function. This way we only need to define the changes and omit everything that is similar.

These two \code{ssp} objects are then passed on to the \code{gridplot} function. Here we make a $2\times 2$ grid, of which the bottom row is for the legends, but the function can also automatically determine the number of rows and columns and the positions of the legends.

\begin{knitrout}
\definecolor{shadecolor}{rgb}{0.969, 0.969, 0.969}\color{fgcolor}\begin{kframe}
\begin{alltt}
\hlstd{ssp_f} \hlkwb{<-} \hlkwd{ssp}\hlstd{(}\hlkwd{list}\hlstd{(marr_seq[biofam3c}\hlopt{$}\hlstd{covariates}\hlopt{$}\hlstd{sex} \hlopt{==} \hlstr{"woman"}\hlstd{,],}
    \hlstd{child_seq[biofam3c}\hlopt{$}\hlstd{covariates}\hlopt{$}\hlstd{sex} \hlopt{==} \hlstr{"woman"}\hlstd{,],}
    \hlstd{left_seq[biofam3c}\hlopt{$}\hlstd{covariates}\hlopt{$}\hlstd{sex} \hlopt{==} \hlstr{"woman"}\hlstd{,]),}
  \hlkwc{type} \hlstd{=} \hlstr{"I"}\hlstd{,} \hlkwc{sortv} \hlstd{=} \hlstr{"mds.obs"}\hlstd{,} \hlkwc{with.legend} \hlstd{=} \hlnum{FALSE}\hlstd{,} \hlkwc{title} \hlstd{=} \hlstr{"Women"}\hlstd{,}
  \hlkwc{ylab.pos} \hlstd{=} \hlkwd{c}\hlstd{(}\hlnum{1}\hlstd{,} \hlnum{2}\hlstd{,} \hlnum{1}\hlstd{),} \hlkwc{xtlab} \hlstd{=} \hlnum{15}\hlopt{:}\hlnum{30}\hlstd{,} \hlkwc{ylab} \hlstd{=} \hlkwd{c}\hlstd{(}\hlstr{"Married"}\hlstd{,} \hlstr{"Children"}\hlstd{,}
    \hlstr{"Residence"}\hlstd{))}

\hlstd{ssp_m} \hlkwb{<-} \hlkwd{update}\hlstd{(ssp_f,} \hlkwc{title} \hlstd{=} \hlstr{"Men"}\hlstd{,}
  \hlkwc{x} \hlstd{=} \hlkwd{list}\hlstd{(marr_seq[biofam3c}\hlopt{$}\hlstd{covariates}\hlopt{$}\hlstd{sex} \hlopt{==} \hlstr{"man"}\hlstd{,],}
    \hlstd{child_seq[biofam3c}\hlopt{$}\hlstd{covariates}\hlopt{$}\hlstd{sex} \hlopt{==} \hlstr{"man"}\hlstd{,],}
    \hlstd{left_seq[biofam3c}\hlopt{$}\hlstd{covariates}\hlopt{$}\hlstd{sex} \hlopt{==} \hlstr{"man"}\hlstd{,]))}

\hlkwd{gridplot}\hlstd{(}\hlkwd{list}\hlstd{(ssp_f, ssp_m),} \hlkwc{ncol} \hlstd{=} \hlnum{2}\hlstd{,} \hlkwc{nrow} \hlstd{=} \hlnum{2}\hlstd{,} \hlkwc{byrow} \hlstd{=} \hlnum{TRUE}\hlstd{,}
  \hlkwc{legend.pos} \hlstd{=} \hlstr{"bottom"}\hlstd{,} \hlkwc{legend.pos2} \hlstd{=} \hlstr{"top"}\hlstd{,} \hlkwc{row.prop} \hlstd{=} \hlkwd{c}\hlstd{(}\hlnum{0.65}\hlstd{,} \hlnum{0.35}\hlstd{))}
\end{alltt}
\end{kframe}
\end{knitrout}

For more examples on visualization, see a supplementary vignette on CRAN \citep{Helske2017a}.

\newpage
\subsection{Hidden Markov models}
\label{sec:exHMM}

We start by showing how to fit an HMM for single-channel \code{biofam} data. The model is initialized with the \code{build\_hmm} function which creates an object of class \code{hmm}. The simplest way is to use automatic starting values by giving the number of hidden states.
\begin{knitrout}
\definecolor{shadecolor}{rgb}{0.969, 0.969, 0.969}\color{fgcolor}\begin{kframe}
\begin{alltt}
\hlstd{sc_initmod_random} \hlkwb{<-} \hlkwd{build_hmm}\hlstd{(}\hlkwc{observations} \hlstd{= biofam_seq,} \hlkwc{n_states} \hlstd{=} \hlnum{5}\hlstd{)}
\end{alltt}
\end{kframe}
\end{knitrout}

It is, however, often advisable to set starting values for initial, transition, and emission probabilities manually. Here the hidden states are regarded as more general life stages, during which individuals are more likely to meet certain observable life events. We expect that the life stages are somehow related to age, so constructing starting values from the observed state frequencies by age group seems like an option worth a try (these are easily computed using the \code{seqstatf} function in \pkg{TraMineR}). We construct a model with four hidden states using age groups 15--18, 19--21, 22--24, 25--27 and 28--30.

The \code{fit\_model} function uses the probabilities given by the initial model as starting values when estimating the parameters. Only positive probabilities are estimated; zero values are fixed to zero. Thus, the amount of 0.1 is added to each value in case of zero-frequencies in some categories (at this point we do not want to fix any parameters to zero). Each row is divided by its sum, so that the row sums equal to 1. 

\begin{knitrout}
\definecolor{shadecolor}{rgb}{0.969, 0.969, 0.969}\color{fgcolor}\begin{kframe}
\begin{alltt}
\hlstd{sc_init} \hlkwb{<-} \hlkwd{c}\hlstd{(}\hlnum{0.9}\hlstd{,} \hlnum{0.06}\hlstd{,} \hlnum{0.02}\hlstd{,} \hlnum{0.01}\hlstd{,} \hlnum{0.01}\hlstd{)}

\hlstd{sc_trans} \hlkwb{<-} \hlkwd{matrix}\hlstd{(}\hlkwd{c}\hlstd{(}\hlnum{0.80}\hlstd{,} \hlnum{0.10}\hlstd{,} \hlnum{0.05}\hlstd{,} \hlnum{0.03}\hlstd{,} \hlnum{0.02}\hlstd{,} \hlnum{0.02}\hlstd{,} \hlnum{0.80}\hlstd{,} \hlnum{0.10}\hlstd{,}
  \hlnum{0.05}\hlstd{,} \hlnum{0.03}\hlstd{,} \hlnum{0.02}\hlstd{,} \hlnum{0.03}\hlstd{,} \hlnum{0.80}\hlstd{,} \hlnum{0.10}\hlstd{,} \hlnum{0.05}\hlstd{,} \hlnum{0.02}\hlstd{,} \hlnum{0.03}\hlstd{,} \hlnum{0.05}\hlstd{,} \hlnum{0.80}\hlstd{,} \hlnum{0.10}\hlstd{,}
  \hlnum{0.02}\hlstd{,} \hlnum{0.03}\hlstd{,} \hlnum{0.05}\hlstd{,} \hlnum{0.05}\hlstd{,} \hlnum{0.85}\hlstd{),} \hlkwc{nrow} \hlstd{=} \hlnum{5}\hlstd{,} \hlkwc{ncol} \hlstd{=} \hlnum{5}\hlstd{,} \hlkwc{byrow} \hlstd{=} \hlnum{TRUE}\hlstd{)}

\hlstd{sc_emiss} \hlkwb{<-} \hlkwd{matrix}\hlstd{(}\hlnum{NA}\hlstd{,} \hlkwc{nrow} \hlstd{=} \hlnum{5}\hlstd{,} \hlkwc{ncol} \hlstd{=} \hlnum{8}\hlstd{)}
\hlstd{sc_emiss[}\hlnum{1}\hlstd{,]} \hlkwb{<-} \hlkwd{seqstatf}\hlstd{(biofam_seq[,} \hlnum{1}\hlopt{:}\hlnum{4}\hlstd{])[,} \hlnum{2}\hlstd{]} \hlopt{+} \hlnum{0.1}
\hlstd{sc_emiss[}\hlnum{2}\hlstd{,]} \hlkwb{<-} \hlkwd{seqstatf}\hlstd{(biofam_seq[,} \hlnum{5}\hlopt{:}\hlnum{7}\hlstd{])[,} \hlnum{2}\hlstd{]} \hlopt{+} \hlnum{0.1}
\hlstd{sc_emiss[}\hlnum{3}\hlstd{,]} \hlkwb{<-} \hlkwd{seqstatf}\hlstd{(biofam_seq[,} \hlnum{8}\hlopt{:}\hlnum{10}\hlstd{])[,} \hlnum{2}\hlstd{]} \hlopt{+} \hlnum{0.1}
\hlstd{sc_emiss[}\hlnum{4}\hlstd{,]} \hlkwb{<-} \hlkwd{seqstatf}\hlstd{(biofam_seq[,} \hlnum{11}\hlopt{:}\hlnum{13}\hlstd{])[,} \hlnum{2}\hlstd{]} \hlopt{+} \hlnum{0.1}
\hlstd{sc_emiss[}\hlnum{5}\hlstd{,]} \hlkwb{<-} \hlkwd{seqstatf}\hlstd{(biofam_seq[,} \hlnum{14}\hlopt{:}\hlnum{16}\hlstd{])[,} \hlnum{2}\hlstd{]} \hlopt{+} \hlnum{0.1}
\hlstd{sc_emiss} \hlkwb{<-} \hlstd{sc_emiss} \hlopt{/} \hlkwd{rowSums}\hlstd{(sc_emiss)}

\hlkwd{rownames}\hlstd{(sc_trans)} \hlkwb{<-} \hlkwd{colnames}\hlstd{(sc_trans)} \hlkwb{<-} \hlkwd{rownames}\hlstd{(sc_emiss)} \hlkwb{<-}
  \hlkwd{paste}\hlstd{(}\hlstr{"State"}\hlstd{,} \hlnum{1}\hlopt{:}\hlnum{5}\hlstd{)}

\hlkwd{colnames}\hlstd{(sc_emiss)} \hlkwb{<-} \hlkwd{attr}\hlstd{(biofam_seq,} \hlstr{"labels"}\hlstd{)}

\hlstd{sc_trans}
\end{alltt}
\begin{verbatim}
##         State 1 State 2 State 3 State 4 State 5
## State 1    0.80    0.10    0.05    0.03    0.02
## State 2    0.02    0.80    0.10    0.05    0.03
## State 3    0.02    0.03    0.80    0.10    0.05
## State 4    0.02    0.03    0.05    0.80    0.10
## State 5    0.02    0.03    0.05    0.05    0.85
\end{verbatim}
\begin{alltt}
\hlkwd{round}\hlstd{(sc_emiss,} \hlnum{3}\hlstd{)}
\end{alltt}
\begin{verbatim}
##         parent  left married left+marr child left+child left+marr+ch
## State 1  0.928 0.063   0.002     0.002 0.001      0.001        0.002
## State 2  0.701 0.218   0.018     0.028 0.001      0.004        0.029
## State 3  0.417 0.290   0.050     0.114 0.001      0.006        0.117
## State 4  0.204 0.231   0.080     0.201 0.002      0.009        0.256
## State 5  0.101 0.157   0.097     0.196 0.002      0.013        0.400
##         divorced
## State 1    0.001
## State 2    0.001
## State 3    0.005
## State 4    0.018
## State 5    0.034
\end{verbatim}
\end{kframe}
\end{knitrout}
Now, the \code{build\_hmm} checks that the data and matrices are of the right form.
\begin{knitrout}
\definecolor{shadecolor}{rgb}{0.969, 0.969, 0.969}\color{fgcolor}\begin{kframe}
\begin{alltt}
\hlstd{sc_initmod} \hlkwb{<-} \hlkwd{build_hmm}\hlstd{(}\hlkwc{observations} \hlstd{= biofam_seq,} \hlkwc{initial_probs} \hlstd{= sc_init,}
  \hlkwc{transition_probs} \hlstd{= sc_trans,} \hlkwc{emission_probs} \hlstd{= sc_emiss)}
\end{alltt}
\end{kframe}
\end{knitrout}
We then use the \code{fit\_model} function for parameter estimation. Here we estimate the model using the default options of the EM step.
\begin{knitrout}
\definecolor{shadecolor}{rgb}{0.969, 0.969, 0.969}\color{fgcolor}\begin{kframe}
\begin{alltt}
\hlstd{sc_fit} \hlkwb{<-} \hlkwd{fit_model}\hlstd{(sc_initmod)}
\end{alltt}
\end{kframe}
\end{knitrout}
The fitting function returns the estimated model, its log-likelihood, and information on the optimization steps.
\begin{knitrout}
\definecolor{shadecolor}{rgb}{0.969, 0.969, 0.969}\color{fgcolor}\begin{kframe}
\begin{alltt}
\hlstd{sc_fit}\hlopt{$}\hlstd{logLik}
\end{alltt}
\begin{verbatim}
## [1] -16781.99
\end{verbatim}
\end{kframe}
\end{knitrout}

\begin{knitrout}
\definecolor{shadecolor}{rgb}{0.969, 0.969, 0.969}\color{fgcolor}\begin{kframe}
\begin{alltt}
\hlstd{sc_fit}\hlopt{$}\hlstd{model}
\end{alltt}
\begin{verbatim}
## Initial probabilities :
## State 1 State 2 State 3 State 4 State 5 
##   0.986   0.000   0.014   0.000   0.000 
## 
## Transition probabilities :
##          to
## from      State 1 State 2 State 3 State 4 State 5
##   State 1   0.786   0.175  0.0391 0.00000  0.0000
##   State 2   0.000   0.786  0.0751 0.07568  0.0631
##   State 3   0.000   0.000  0.8898 0.08342  0.0267
##   State 4   0.000   0.000  0.0000 0.78738  0.2126
##   State 5   0.000   0.000  0.0000 0.00136  0.9986
## 
## Emission probabilities :
##            symbol_names
## state_names 0 1       2     3       4      5     6      7
##     State 1 1 0 0.00000 0.000 0.00000 0.0000 0.000 0.0000
##     State 2 1 0 0.00000 0.000 0.00000 0.0000 0.000 0.0000
##     State 3 0 1 0.00000 0.000 0.00000 0.0000 0.000 0.0000
##     State 4 0 0 0.00195 0.992 0.00581 0.0000 0.000 0.0000
##     State 5 0 0 0.21508 0.000 0.00000 0.0246 0.713 0.0474
\end{verbatim}
\end{kframe}
\end{knitrout}

As a multichannel example we fit a 5-state model for the 3-channel data. Emission probabilities are now given as a list of three emission matrices, one for each channel. The \code{alphabet} function from the \pkg{TraMineR} package can be used to check the order of the observed states -- the same order is used in the \code{build} functions. Here we construct a left-to-right model where transitions to earlier states are not allowed, so the transition matrix is upper-triangular. This seems like a valid option from a life-course perspective. Also, in the previous single-channel model of the same data the transition matrix was estimated almost upper triangular. We also give names for channels -- these are used when printing and plotting the model.

We estimate model parameters using the local step with the default L-BFGS algorithm using parallel computation with 4 threads.

\begin{knitrout}
\definecolor{shadecolor}{rgb}{0.969, 0.969, 0.969}\color{fgcolor}\begin{kframe}
\begin{alltt}
\hlstd{mc_init} \hlkwb{<-} \hlkwd{c}\hlstd{(}\hlnum{0.9}\hlstd{,} \hlnum{0.05}\hlstd{,} \hlnum{0.02}\hlstd{,} \hlnum{0.02}\hlstd{,} \hlnum{0.01}\hlstd{)}

\hlstd{mc_trans} \hlkwb{<-} \hlkwd{matrix}\hlstd{(}\hlkwd{c}\hlstd{(}\hlnum{0.80}\hlstd{,} \hlnum{0.10}\hlstd{,} \hlnum{0.05}\hlstd{,} \hlnum{0.03}\hlstd{,} \hlnum{0.02}\hlstd{,} \hlnum{0}\hlstd{,} \hlnum{0.90}\hlstd{,} \hlnum{0.05}\hlstd{,} \hlnum{0.03}\hlstd{,}
  \hlnum{0.02}\hlstd{,} \hlnum{0}\hlstd{,} \hlnum{0}\hlstd{,} \hlnum{0.90}\hlstd{,} \hlnum{0.07}\hlstd{,} \hlnum{0.03}\hlstd{,} \hlnum{0}\hlstd{,} \hlnum{0}\hlstd{,} \hlnum{0}\hlstd{,} \hlnum{0.90}\hlstd{,} \hlnum{0.10}\hlstd{,} \hlnum{0}\hlstd{,} \hlnum{0}\hlstd{,} \hlnum{0}\hlstd{,} \hlnum{0}\hlstd{,} \hlnum{1}\hlstd{),}
  \hlkwc{nrow} \hlstd{=} \hlnum{5}\hlstd{,} \hlkwc{ncol} \hlstd{=} \hlnum{5}\hlstd{,} \hlkwc{byrow} \hlstd{=} \hlnum{TRUE}\hlstd{)}

\hlstd{mc_emiss_marr} \hlkwb{<-} \hlkwd{matrix}\hlstd{(}\hlkwd{c}\hlstd{(}\hlnum{0.90}\hlstd{,} \hlnum{0.05}\hlstd{,} \hlnum{0.05}\hlstd{,} \hlnum{0.90}\hlstd{,} \hlnum{0.05}\hlstd{,} \hlnum{0.05}\hlstd{,} \hlnum{0.05}\hlstd{,} \hlnum{0.90}\hlstd{,}
  \hlnum{0.05}\hlstd{,} \hlnum{0.05}\hlstd{,} \hlnum{0.90}\hlstd{,} \hlnum{0.05}\hlstd{,} \hlnum{0.30}\hlstd{,} \hlnum{0.30}\hlstd{,} \hlnum{0.40}\hlstd{),} \hlkwc{nrow} \hlstd{=} \hlnum{5}\hlstd{,} \hlkwc{ncol} \hlstd{=} \hlnum{3}\hlstd{,}
  \hlkwc{byrow} \hlstd{=} \hlnum{TRUE}\hlstd{)}

\hlstd{mc_emiss_child} \hlkwb{<-} \hlkwd{matrix}\hlstd{(}\hlkwd{c}\hlstd{(}\hlnum{0.9}\hlstd{,} \hlnum{0.1}\hlstd{,} \hlnum{0.9}\hlstd{,} \hlnum{0.1}\hlstd{,} \hlnum{0.1}\hlstd{,} \hlnum{0.9}\hlstd{,} \hlnum{0.1}\hlstd{,} \hlnum{0.9}\hlstd{,} \hlnum{0.5}\hlstd{,}
  \hlnum{0.5}\hlstd{),} \hlkwc{nrow} \hlstd{=} \hlnum{5}\hlstd{,} \hlkwc{ncol} \hlstd{=} \hlnum{2}\hlstd{,} \hlkwc{byrow} \hlstd{=} \hlnum{TRUE}\hlstd{)}

\hlstd{mc_emiss_left} \hlkwb{<-} \hlkwd{matrix}\hlstd{(}\hlkwd{c}\hlstd{(}\hlnum{0.9}\hlstd{,} \hlnum{0.1}\hlstd{,} \hlnum{0.1}\hlstd{,} \hlnum{0.9}\hlstd{,} \hlnum{0.1}\hlstd{,} \hlnum{0.9}\hlstd{,} \hlnum{0.1}\hlstd{,} \hlnum{0.9}\hlstd{,} \hlnum{0.5}\hlstd{,}
  \hlnum{0.5}\hlstd{),} \hlkwc{nrow} \hlstd{=} \hlnum{5}\hlstd{,} \hlkwc{ncol} \hlstd{=} \hlnum{2}\hlstd{,} \hlkwc{byrow} \hlstd{=} \hlnum{TRUE}\hlstd{)}

\hlstd{mc_obs} \hlkwb{<-} \hlkwd{list}\hlstd{(marr_seq, child_seq, left_seq)}

\hlstd{mc_emiss} \hlkwb{<-} \hlkwd{list}\hlstd{(mc_emiss_marr, mc_emiss_child, mc_emiss_left)}

\hlstd{mc_initmod} \hlkwb{<-} \hlkwd{build_hmm}\hlstd{(}\hlkwc{observations} \hlstd{= mc_obs,} \hlkwc{initial_probs} \hlstd{= mc_init,}
  \hlkwc{transition_probs} \hlstd{= mc_trans,} \hlkwc{emission_probs} \hlstd{= mc_emiss,}
  \hlkwc{channel_names} \hlstd{=} \hlkwd{c}\hlstd{(}\hlstr{"Marriage"}\hlstd{,} \hlstr{"Parenthood"}\hlstd{,} \hlstr{"Residence"}\hlstd{))}

\hlcom{# For CRAN vignette: load the estimated model object for speed-up}
\hlkwd{data}\hlstd{(}\hlstr{"hmm_biofam"}\hlstd{)}
\hlcom{# mc_fit <- fit_model(mc_initmod, em_step = FALSE, local_step = TRUE,}
\hlcom{# threads = 4)}
\end{alltt}
\end{kframe}
\end{knitrout}

We store the model as a separate object for the ease of use and then compute BIC.

\begin{knitrout}
\definecolor{shadecolor}{rgb}{0.969, 0.969, 0.969}\color{fgcolor}\begin{kframe}
\begin{alltt}
\hlcom{# Vignette: already loaded hmm_biofam}
\hlcom{# hmm_biofam <- mc_fit$model}
\hlkwd{BIC}\hlstd{(hmm_biofam)}
\end{alltt}
\begin{verbatim}
## [1] 28842.7
\end{verbatim}
\end{kframe}
\end{knitrout}

\subsection{Clustering and mixture hidden Markov models}
\label{sec:code_MHMM}

When fitting mixture hidden Markov models, the starting values are given as lists, with one component per cluster. For multichannel data, emission probabilities are given as a list of lists. Here we fit a model for two clusters with 5 and 4 hidden states. For the cluster with five states we use the same starting values as for the multichannel HMM described earlier. Covariates are defined with the usual \code{formula} and \code{data} arguments. Here we use sex and birth cohort to explain cluster memberships.

We fit a model using 100 random restarts of the EM algorithm followed by the local L-BFGS method. Again we use parallel computation.

\begin{knitrout}
\definecolor{shadecolor}{rgb}{0.969, 0.969, 0.969}\color{fgcolor}\begin{kframe}
\begin{alltt}
\hlstd{mc_init2} \hlkwb{<-} \hlkwd{c}\hlstd{(}\hlnum{0.9}\hlstd{,} \hlnum{0.05}\hlstd{,} \hlnum{0.03}\hlstd{,} \hlnum{0.02}\hlstd{)}

\hlstd{mc_trans2} \hlkwb{<-} \hlkwd{matrix}\hlstd{(}\hlkwd{c}\hlstd{(}\hlnum{0.85}\hlstd{,} \hlnum{0.05}\hlstd{,} \hlnum{0.05}\hlstd{,} \hlnum{0.05}\hlstd{,} \hlnum{0}\hlstd{,} \hlnum{0.90}\hlstd{,} \hlnum{0.05}\hlstd{,} \hlnum{0.05}\hlstd{,} \hlnum{0}\hlstd{,} \hlnum{0}\hlstd{,}
  \hlnum{0.95}\hlstd{,} \hlnum{0.05}\hlstd{,} \hlnum{0}\hlstd{,} \hlnum{0}\hlstd{,} \hlnum{0}\hlstd{,} \hlnum{1}\hlstd{),} \hlkwc{nrow} \hlstd{=} \hlnum{4}\hlstd{,} \hlkwc{ncol} \hlstd{=} \hlnum{4}\hlstd{,} \hlkwc{byrow} \hlstd{=} \hlnum{TRUE}\hlstd{)}

\hlstd{mc_emiss_marr2} \hlkwb{<-} \hlkwd{matrix}\hlstd{(}\hlkwd{c}\hlstd{(}\hlnum{0.90}\hlstd{,} \hlnum{0.05}\hlstd{,} \hlnum{0.05}\hlstd{,} \hlnum{0.90}\hlstd{,} \hlnum{0.05}\hlstd{,} \hlnum{0.05}\hlstd{,} \hlnum{0.05}\hlstd{,}
  \hlnum{0.85}\hlstd{,} \hlnum{0.10}\hlstd{,} \hlnum{0.05}\hlstd{,} \hlnum{0.80}\hlstd{,} \hlnum{0.15}\hlstd{),} \hlkwc{nrow} \hlstd{=} \hlnum{4}\hlstd{,} \hlkwc{ncol} \hlstd{=} \hlnum{3}\hlstd{,} \hlkwc{byrow} \hlstd{=} \hlnum{TRUE}\hlstd{)}

\hlstd{mc_emiss_child2} \hlkwb{<-} \hlkwd{matrix}\hlstd{(}\hlkwd{c}\hlstd{(}\hlnum{0.9}\hlstd{,} \hlnum{0.1}\hlstd{,} \hlnum{0.5}\hlstd{,} \hlnum{0.5}\hlstd{,} \hlnum{0.5}\hlstd{,} \hlnum{0.5}\hlstd{,} \hlnum{0.5}\hlstd{,} \hlnum{0.5}\hlstd{),}
  \hlkwc{nrow} \hlstd{=} \hlnum{4}\hlstd{,} \hlkwc{ncol} \hlstd{=} \hlnum{2}\hlstd{,} \hlkwc{byrow} \hlstd{=} \hlnum{TRUE}\hlstd{)}

\hlstd{mc_emiss_left2} \hlkwb{<-} \hlkwd{matrix}\hlstd{(}\hlkwd{c}\hlstd{(}\hlnum{0.9}\hlstd{,} \hlnum{0.1}\hlstd{,} \hlnum{0.5}\hlstd{,} \hlnum{0.5}\hlstd{,} \hlnum{0.5}\hlstd{,} \hlnum{0.5}\hlstd{,} \hlnum{0.5}\hlstd{,} \hlnum{0.5}\hlstd{),}
  \hlkwc{nrow} \hlstd{=} \hlnum{4}\hlstd{,} \hlkwc{ncol} \hlstd{=} \hlnum{2}\hlstd{,} \hlkwc{byrow} \hlstd{=} \hlnum{TRUE}\hlstd{)}

\hlstd{mhmm_init} \hlkwb{<-} \hlkwd{list}\hlstd{(mc_init, mc_init2)}

\hlstd{mhmm_trans} \hlkwb{<-} \hlkwd{list}\hlstd{(mc_trans, mc_trans2)}

\hlstd{mhmm_emiss} \hlkwb{<-} \hlkwd{list}\hlstd{(}\hlkwd{list}\hlstd{(mc_emiss_marr, mc_emiss_child, mc_emiss_left),}
  \hlkwd{list}\hlstd{(mc_emiss_marr2, mc_emiss_child2, mc_emiss_left2))}

\hlstd{biofam3c}\hlopt{$}\hlstd{covariates}\hlopt{$}\hlstd{cohort} \hlkwb{<-} \hlkwd{cut}\hlstd{(biofam3c}\hlopt{$}\hlstd{covariates}\hlopt{$}\hlstd{birthyr,}
  \hlkwd{c}\hlstd{(}\hlnum{1908}\hlstd{,} \hlnum{1935}\hlstd{,} \hlnum{1945}\hlstd{,} \hlnum{1957}\hlstd{))}
\hlstd{biofam3c}\hlopt{$}\hlstd{covariates}\hlopt{$}\hlstd{cohort} \hlkwb{<-} \hlkwd{factor}\hlstd{(biofam3c}\hlopt{$}\hlstd{covariates}\hlopt{$}\hlstd{cohort,}
  \hlkwc{labels}\hlstd{=}\hlkwd{c}\hlstd{(}\hlstr{"1909-1935"}\hlstd{,} \hlstr{"1936-1945"}\hlstd{,} \hlstr{"1946-1957"}\hlstd{))}

\hlstd{init_mhmm} \hlkwb{<-} \hlkwd{build_mhmm}\hlstd{(}\hlkwc{observations} \hlstd{= mc_obs,} \hlkwc{initial_probs} \hlstd{= mhmm_init,}
  \hlkwc{transition_probs} \hlstd{= mhmm_trans,} \hlkwc{emission_probs} \hlstd{= mhmm_emiss,}
  \hlkwc{formula} \hlstd{=} \hlopt{~}\hlstd{sex} \hlopt{+} \hlstd{cohort,} \hlkwc{data} \hlstd{= biofam3c}\hlopt{$}\hlstd{covariates,}
  \hlkwc{channel_names} \hlstd{=} \hlkwd{c}\hlstd{(}\hlstr{"Marriage"}\hlstd{,} \hlstr{"Parenthood"}\hlstd{,} \hlstr{"Residence"}\hlstd{),}
  \hlkwc{cluster_names} \hlstd{=} \hlkwd{c}\hlstd{(}\hlstr{"Cluster 1"}\hlstd{,} \hlstr{"Cluster 2"}\hlstd{))}

\hlcom{# vignette: less restarts and no parallelization}
\hlkwd{set.seed}\hlstd{(}\hlnum{1011}\hlstd{)}
\hlstd{mhmm_fit} \hlkwb{<-} \hlkwd{fit_model}\hlstd{(init_mhmm,} \hlkwc{local_step} \hlstd{=} \hlnum{TRUE}\hlstd{,} \hlkwc{threads} \hlstd{=} \hlnum{1}\hlstd{,}
  \hlkwc{control_em} \hlstd{=} \hlkwd{list}\hlstd{(}\hlkwc{restart} \hlstd{=} \hlkwd{list}\hlstd{(}\hlkwc{times} \hlstd{=} \hlnum{10}\hlstd{)))}
\hlstd{mhmm} \hlkwb{<-} \hlstd{mhmm_fit}\hlopt{$}\hlstd{model}
\end{alltt}
\end{kframe}
\end{knitrout}

The \code{summary} method automatically computes some features for an MHMM, e.g., standard errors for covariates and prior and posterior cluster probabilities for subjects. A \code{print} method shows some summaries of these: estimates and standard errors for covariates (see Section \ref{sec:MHMM}), log-likelihood and BIC, and information on most probable clusters and prior probabilities. Parameter estimates for transitions, emissions, and initial probabilities are omitted by default. The classification table shows mean probabilities of belonging to each cluster by the most probable cluster (defined from posterior cluster probabilities). A good model should have values close to 1 on the diagonal.

\begin{knitrout}
\definecolor{shadecolor}{rgb}{0.969, 0.969, 0.969}\color{fgcolor}\begin{kframe}
\begin{alltt}
\hlkwd{summary}\hlstd{(mhmm,} \hlkwc{conditional_se} \hlstd{=} \hlnum{FALSE}\hlstd{)}
\end{alltt}
\begin{verbatim}
## Covariate effects :
## Cluster 1 is the reference.
## 
## Cluster 2 :
##                  Estimate  Std. error
## (Intercept)        -1.209       0.138
## sexwoman            0.213       0.141
## cohort1936-1945    -0.785       0.172
## cohort1946-1957    -1.238       0.165
## 
## Log-likelihood: -12969.57   BIC: 26592.66 
## 
## Means of prior cluster probabilities :
## Cluster 1 Cluster 2 
##     0.857     0.143 
## 
## Most probable clusters :
##             Cluster 1  Cluster 2
## count            1753        247
## proportion      0.876      0.124
## 
## Classification table :
## Mean cluster probabilities (in columns) by the most probable cluster (rows)
## 
##           Cluster 1 Cluster 2
## Cluster 1    0.9775    0.0225
## Cluster 2    0.0013    0.9987
\end{verbatim}
\end{kframe}
\end{knitrout}

\subsection{Visualizing hidden Markov models}

The figures in Section \ref{seq:vizHMM} illustrate the five-state multichannel HMM fitted in Section \ref{sec:exHMM}.

A basic HMM graph is easily called with the \code{plot} method. Figure \ref{fig:code_plottingHMMbasic} illustrates the default plot.

\begin{knitrout}
\definecolor{shadecolor}{rgb}{0.969, 0.969, 0.969}\color{fgcolor}\begin{kframe}
\begin{alltt}
\hlkwd{plot}\hlstd{(hmm_biofam)}
\end{alltt}
\end{kframe}\begin{figure}

{\centering \includegraphics[width=\linewidth]{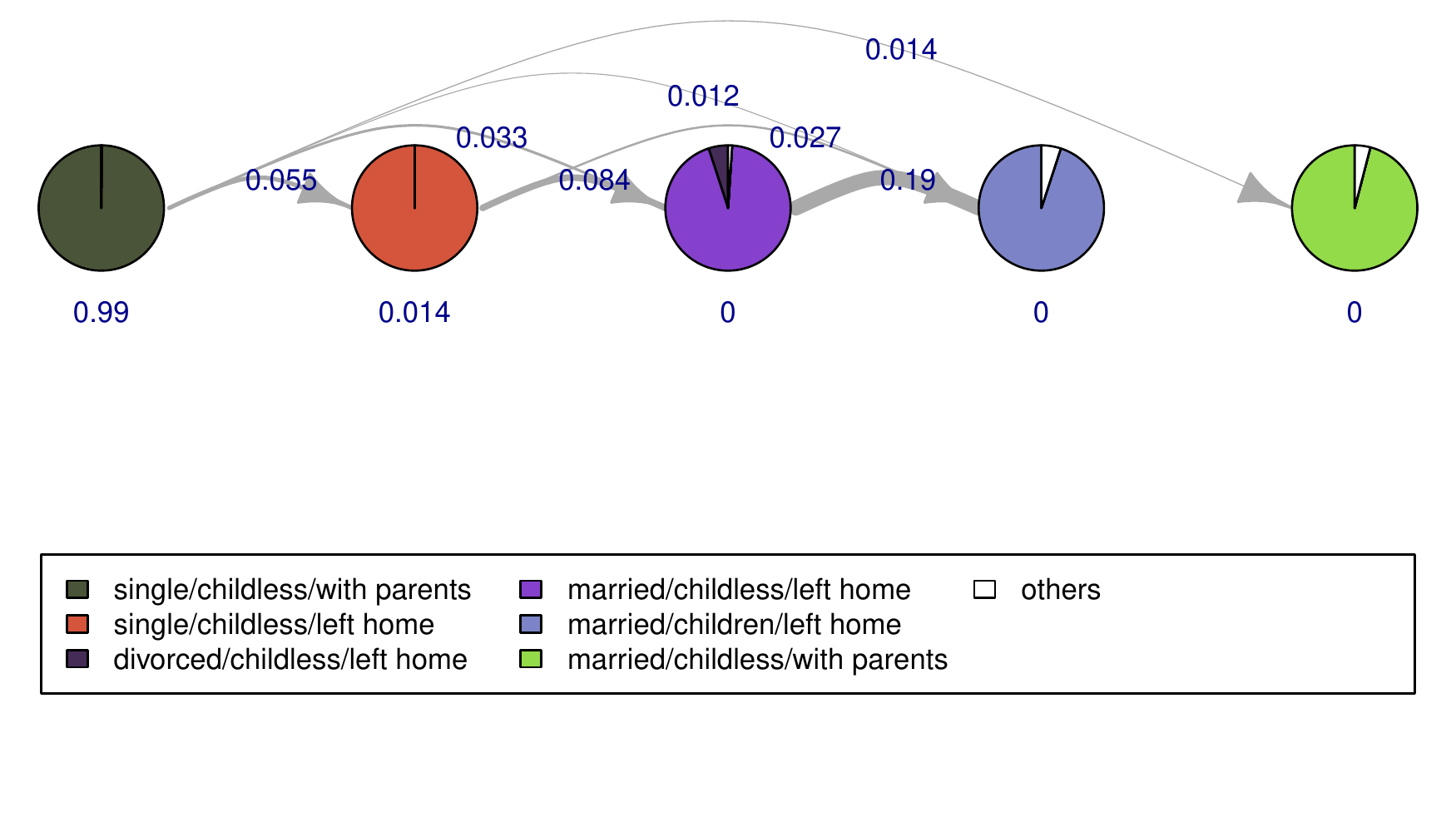} 

}

\caption[A default plot of a hidden Markov model]{A default plot of a hidden Markov model.}\label{fig:code_plottingHMMbasic}
\end{figure}

\end{knitrout}

A simple default plot is a convenient way of visualizing the models during the analysis process, but for publishing it is often better to modify the plot to get an output that best illustrates the structure of the model in hand. Figure \ref{fig:plottingHMM} and Figure \ref{fig:graphicalillustrations5} show two variants of the same model.

\hypertarget{code_plottingHMM}{}\subsubsection[Figure 4]{Figure \ref{fig:plottingHMM}: HMM plot with modifications}

In Figure \ref{fig:plottingHMM} we draw larger vertices, control the distances of initial probabilities (vertex labels), set the curvatures of the edges, give a more descriptive label for the combined slices and give less space for the legend.

\begin{knitrout}
\definecolor{shadecolor}{rgb}{0.969, 0.969, 0.969}\color{fgcolor}\begin{kframe}
\begin{alltt}
\hlkwd{plot}\hlstd{(hmm_biofam,} \hlkwc{vertex.size} \hlstd{=} \hlnum{50}\hlstd{,} \hlkwc{vertex.label.dist} \hlstd{=} \hlnum{1.5}\hlstd{,}
  \hlkwc{edge.curved} \hlstd{=} \hlkwd{c}\hlstd{(}\hlnum{0}\hlstd{,} \hlnum{0.6}\hlstd{,} \hlopt{-}\hlnum{0.8}\hlstd{,} \hlnum{0.6}\hlstd{,} \hlnum{0}\hlstd{,} \hlnum{0.6}\hlstd{,} \hlnum{0}\hlstd{),} \hlkwc{legend.prop} \hlstd{=} \hlnum{0.3}\hlstd{,}
  \hlkwc{combined.slice.label} \hlstd{=} \hlstr{"States with prob. < 0.05"}\hlstd{)}
\end{alltt}
\end{kframe}
\end{knitrout}

\hypertarget{code_graphicalillustrations5}{}\subsubsection[Figure 5]{Figure \ref{fig:graphicalillustrations5}: HMM plot with a different layout}

Here we position the vertices using given coordinates. Coordinates are given in a two-column matrix, with x coordinates in the first column and y coordinates in the second. Arguments \code{xlim} and \code{ylim} set the lengths of the axes, and \code{rescale = FALSE} prevents rescaling the coordinates to the $[-1,1]\times[-1,1]$ interval (the default). We modify the positions of initial probabilities, fix edge widths to 1, reduce the size of the arrows in edges, position legend on top of the figure, and print labels in two columns in the legend. Parameter values are shown with one significant digit. All emission probabilities are shown regardless of their value (\code{combine.slices = 0}).

New colors are set from the ready-defined \code{colorpalette} data. The \pkg{seqHMM} package uses these palettes when determining colors automatically, e.g., in the \code{mc\_to\_sc} function. Since here there are 10 combined states, the default color palette is number 10. To get different colors, we choose the ten first colors from palette number 14.

\begin{knitrout}
\definecolor{shadecolor}{rgb}{0.969, 0.969, 0.969}\color{fgcolor}\begin{kframe}
\begin{alltt}
\hlstd{vertex_layout} \hlkwb{<-} \hlkwd{matrix}\hlstd{(}\hlkwd{c}\hlstd{(}\hlnum{1}\hlstd{,} \hlnum{2}\hlstd{,} \hlnum{2}\hlstd{,} \hlnum{3}\hlstd{,} \hlnum{1}\hlstd{,} \hlnum{0}\hlstd{,} \hlnum{0.5}\hlstd{,} \hlopt{-}\hlnum{0.5}\hlstd{,} \hlnum{0}\hlstd{,} \hlopt{-}\hlnum{1}\hlstd{),}
  \hlkwc{ncol} \hlstd{=} \hlnum{2}\hlstd{)}

\hlkwd{plot}\hlstd{(hmm_biofam,} \hlkwc{layout} \hlstd{= vertex_layout,} \hlkwc{xlim} \hlstd{=} \hlkwd{c}\hlstd{(}\hlnum{0.5}\hlstd{,} \hlnum{3.5}\hlstd{),}
  \hlkwc{ylim} \hlstd{=} \hlkwd{c}\hlstd{(}\hlopt{-}\hlnum{1.5}\hlstd{,} \hlnum{1}\hlstd{),} \hlkwc{rescale} \hlstd{=} \hlnum{FALSE}\hlstd{,} \hlkwc{vertex.size} \hlstd{=} \hlnum{50}\hlstd{,}
  \hlkwc{vertex.label.pos} \hlstd{=} \hlkwd{c}\hlstd{(}\hlstr{"left"}\hlstd{,} \hlstr{"top"}\hlstd{,} \hlstr{"bottom"}\hlstd{,} \hlstr{"right"}\hlstd{,} \hlstr{"left"}\hlstd{),}
  \hlkwc{edge.curved} \hlstd{=} \hlnum{FALSE}\hlstd{,} \hlkwc{edge.width} \hlstd{=} \hlnum{1}\hlstd{,} \hlkwc{edge.arrow.size} \hlstd{=} \hlnum{1}\hlstd{,}
  \hlkwc{with.legend} \hlstd{=} \hlstr{"left"}\hlstd{,} \hlkwc{legend.prop} \hlstd{=} \hlnum{0.4}\hlstd{,} \hlkwc{label.signif} \hlstd{=} \hlnum{1}\hlstd{,}
  \hlkwc{combine.slices} \hlstd{=} \hlnum{0}\hlstd{,} \hlkwc{cpal} \hlstd{= colorpalette[[}\hlnum{30}\hlstd{]][}\hlkwd{c}\hlstd{(}\hlnum{14}\hlopt{:}\hlnum{5}\hlstd{)])}
\end{alltt}
\end{kframe}
\end{knitrout}

\hypertarget{ssplotHMM}{}\subsubsection*{Figure \ref{fig:ssplotHMM}: \code{ssplot} for an HMM object}

Plotting observed and hidden state sequences is easy with the \code{ssplot} function: the function accepts an \code{hmm} object instead of (a list of) \code{stslist}s. If hidden state paths are not provided, the function automatically computes them when needed.

\begin{knitrout}
\definecolor{shadecolor}{rgb}{0.969, 0.969, 0.969}\color{fgcolor}\begin{kframe}
\begin{alltt}
\hlkwd{ssplot}\hlstd{(hmm_biofam,} \hlkwc{plots} \hlstd{=} \hlstr{"both"}\hlstd{,} \hlkwc{type} \hlstd{=} \hlstr{"I"}\hlstd{,} \hlkwc{sortv} \hlstd{=} \hlstr{"mds.hidden"}\hlstd{,}
  \hlkwc{title} \hlstd{=} \hlstr{"Observed and hidden state sequences"}\hlstd{,} \hlkwc{xtlab} \hlstd{=} \hlnum{15}\hlopt{:}\hlnum{30}\hlstd{,}
  \hlkwc{xlab} \hlstd{=} \hlstr{"Age"}\hlstd{)}
\end{alltt}
\end{kframe}
\end{knitrout}

\subsection{Visualizing mixture hidden Markov models}

Objects of class \code{mhmm} have similar plotting methods to \code{hmm} objects. The default way of visualizing a model is to plot in an interactive mode, where the model for each cluster is plotted separately. Another option is a combined plot with all models in one plot, although it can be difficult to fit several graphs and legends in one figure.

Figure \ref{fig:code_plottingMHMMbasic} illustrates the MHMM fitted in Section \ref{sec:code_MHMM}. By setting \texttt{interactive = FALSE} and \code{nrow = 2} we plot graphs in a grid with two rows. The rest of the arguments are similar to basic HMM plotting and apply for all the graphs.

\begin{knitrout}
\definecolor{shadecolor}{rgb}{0.969, 0.969, 0.969}\color{fgcolor}\begin{kframe}
\begin{alltt}
\hlkwd{plot}\hlstd{(mhmm,} \hlkwc{interactive} \hlstd{=} \hlnum{FALSE}\hlstd{,} \hlkwc{nrow} \hlstd{=} \hlnum{2}\hlstd{,} \hlkwc{legend.prop} \hlstd{=} \hlnum{0.45}\hlstd{,}
  \hlkwc{vertex.size} \hlstd{=} \hlnum{50}\hlstd{,} \hlkwc{vertex.label.cex} \hlstd{=} \hlnum{1.3}\hlstd{,} \hlkwc{cex.legend} \hlstd{=} \hlnum{1.3}\hlstd{,}
  \hlkwc{edge.curved} \hlstd{=} \hlnum{0.65}\hlstd{,} \hlkwc{edge.label.cex} \hlstd{=} \hlnum{1.3}\hlstd{,} \hlkwc{edge.arrow.size} \hlstd{=} \hlnum{0.8}\hlstd{)}
\end{alltt}
\end{kframe}\begin{figure}

{\centering \includegraphics[width=\maxwidth]{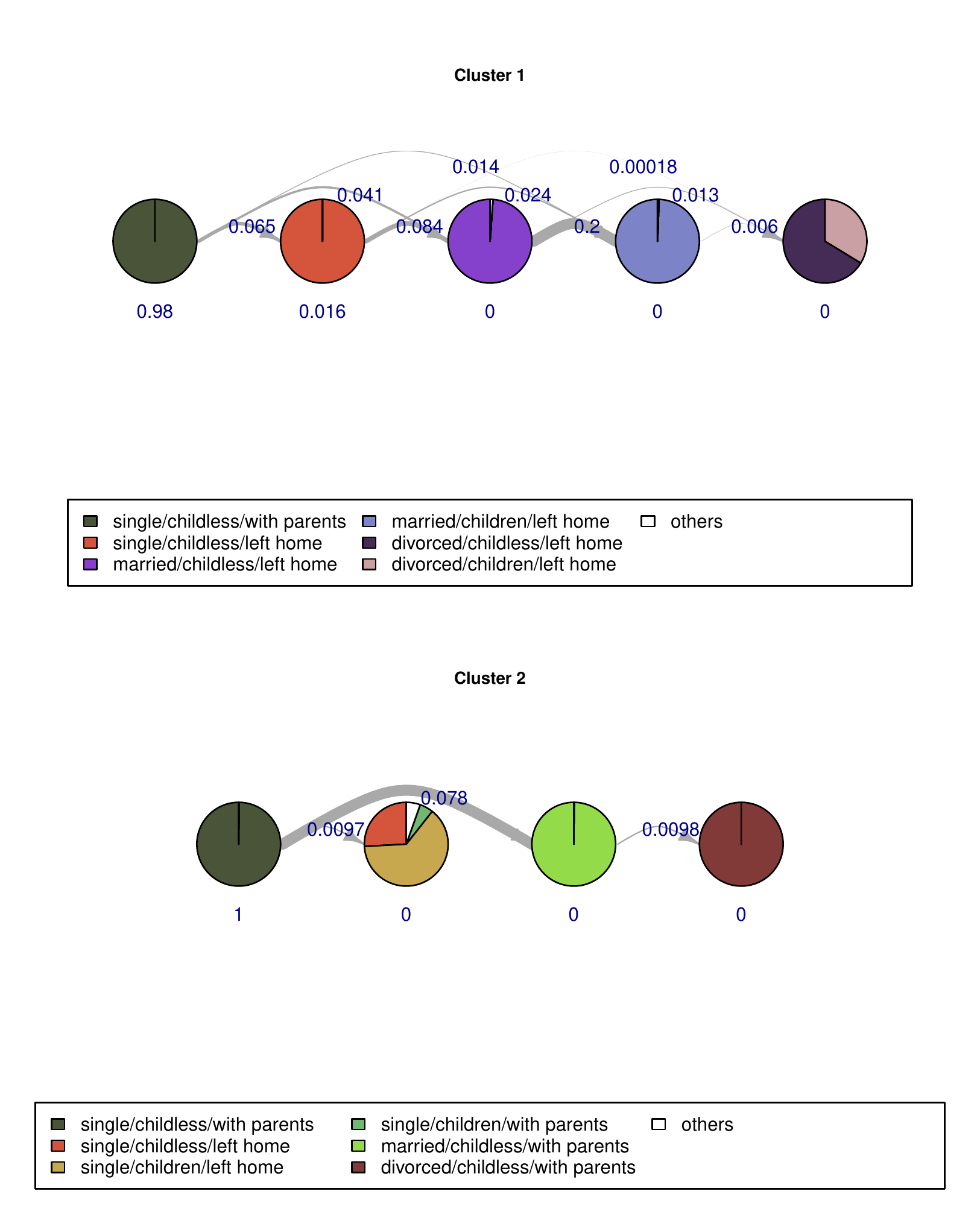} 

}

\caption[Plotting submodels of an MHMM with the \code{plot} method]{Plotting submodels of an MHMM with the \code{plot} method.}\label{fig:code_plottingMHMMbasic}
\end{figure}

\end{knitrout}

The equivalent of the \code{ssplot} function for \code{mhmm} objects is \code{mssplot}. It shows data and/or hidden paths one cluster at a time. The function is interactive if more than one cluster is plotted (thus omitted here). Subjects are allocated to clusters according to the most probable hidden state paths.

\begin{knitrout}
\definecolor{shadecolor}{rgb}{0.969, 0.969, 0.969}\color{fgcolor}\begin{kframe}
\begin{alltt}
\hlkwd{mssplot}\hlstd{(mhmm,} \hlkwc{ask} \hlstd{=} \hlnum{TRUE}\hlstd{)}
\end{alltt}
\end{kframe}
\end{knitrout}

If the user wants more control than the default \code{mhmm} plotting functions offer, they can use the \code{separate\_mhmm} function to convert a \code{mhmm} object into a list of separate \code{hmm} objects. These can then be plotted as any \code{hmm} objects, e.g., use \code{ssp} and \code{gridplot} for plotting sequences and hidden paths of each cluster into the same figure.

\section{Conclusion}

Hidden Markov models are useful in various longitudinal settings with categorical observations. They can be used for accounting measurement error in the observations \citep[e.g., drug use as in][]{Vermunt2008}, for detecting true unobservable states \citep[e.g., different periods of the bipolar disorder as in][]{Lopez2008}, and for compressing information across several types of observations \citep[e.g., finding general life stages as in][]{Helske2016a}.

The \pkg{seqHMM} package is designed for analyzing categorical sequences with hidden Markov models and mixture hidden Markov models, as well as their restricted variants Markov models, mixture Markov models, and latent class models. It can handle many types of data from a single sequence to multiple multichannel sequences. Covariates can be included in MHMMs to explain cluster membership. The package also offers versatile plotting options for sequence data and HMMs, and can easily convert multichannel sequence data and models into single-channel representations.

Parameter estimation in (M)HMMs is often very sensitive to starting values. To deal with that, \pkg{seqHMM} offers several fitting options with global and local optimization using direct numerical estimation and the EM algorithm.

Almost all intensive computations are done in C++. The package also supports parallel computation.

Especially combined with the \pkg{TraMineR} package, \pkg{seqHMM} is designed to offer tools for the whole analysis process from data preparation and description to model fitting, evaluation, and visualization. In future we plan to develop MHMMs to deal with time-varying covariates in transition and emission matrices \citep{Bartolucci2012}, and add an option to incorporate sampling weights for model estimation. Also, the computational efficiency of the restricted variants of (M)HMMs, such as latent class models, could be improved by taking account of the restricted structure of those models in EM and log-likelihood computations.

\section*{Acknowledgements}

Satu Helske is grateful for support for this research from the John Fell Oxford University Press (OUP) Research Fund and the Department of Mathematics and Statistics at the University of Jyv{\"a}skyl{\"a}, Finland, and Jouni Helske for the Emil Aaltonen Foundation and the Academy of Finland (research grants 284513 and 312605).

We also wish to thank Mervi Eerola and Jukka Nyblom as well as the editor and two anonymous referees for their helpful comments and suggestions. Comments, suggestions, and bug reports from various users of \pkg{seqHMM} have also been highly appreciated.

\appendix

\section{Notations}
\begin{tabular}{l p{0.67\textwidth}}
Symbol & Meaning \\
\hline
$Y_i$            & Observation sequences of subject $i, i=1\ldots,N$ \\
$\textbf{y}_{it}$         & Observations of subject $i$ at time $t, t=1,\ldots,T$ \\
$y_{itc}$                 & Observation of subject $i$ at time $t$ in channel $c, c=1,\ldots,C$ \\
$m_c \in\{1,\ldots,M_c\}$ & Observed state space for channel $c$ \\
$z_{it}$                  & Hidden state at time $t$ for subject $i$ \\
$s\in \{1,\ldots,S\}$     & Hidden state space \\
$A=\{a_{sr}\}$            & Transition matrix of size $S\times S$\\
$a_{sr}=P(z_t=r|z_{t-1}=s)$ & Transition probability between hidden states $s$ and $r$ \\
$B_c=\{b_s(m_c)\}$        & Emission matrix of size $S\times M_c$ for channel $c$ \\
$b_s(m_c)=P(y_{itc}=m_c|z_{it}=s)$ & Emission probability of observed state $m_c$ in channel $c$ given hidden state $s$ \\
$b_s(\textbf{y}_{it})=b_s(y_{it1})\cdots b_s(y_{itC})$ & Joint emission probability of observations at time $t$ in channels $1,\ldots,C$ given hidden state $s$ \\
$\pi=(\pi_1,\ldots,\pi_S)^{\top}$ & Vector of initial probabilities \\
$\pi_s=P(z_1=s)$          & Initial probability of hidden state $s$ \\
$\hat{z}_i(Y_i)$ & The most probable hidden state sequence for subject $i$ \\
$\textbf{x}_i$            & Covariates of subject $i$ \\
$\mathcal{M}_k, k=1,\ldots,K$ & Submodel for cluster $k$ (latent class/cluster) \\
$w_{ik}$                  & Probability of cluster $k$ for subject $i$ \\
$\gamma_{k}$               & Regression coefficients for cluster $k$ \\
$\{\pi^k, A^k, B_1^k, \ldots, B_C^k, \gamma_k\}$ & Model parameters for cluster $k$ \\
\end{tabular}


\bibliography{references}

\end{document}